\documentclass[aps,nofootinbib,longbibliography]{revtex4-1}

\usepackage{relsize}
\usepackage[intlimits]{amsmath}
\usepackage{amsmath}
\usepackage{blkarray}
\usepackage{exscale}

\usepackage{graphicx}
\usepackage{dcolumn}
\usepackage{bm}
\usepackage{multirow}                               
\usepackage{subfig}
\usepackage{array,ragged2e}

\usepackage{float}
\usepackage[justification=justified]{caption}
\usepackage{caption}
\usepackage{mathtools}
\usepackage{epstopdf}
\usepackage[toc,page]{appendix}

\usepackage[font={small}]{caption}

\usepackage{multirow}
\usepackage{graphicx}
\usepackage{color}
\usepackage{booktabs}
\usepackage{epstopdf}
\usepackage{color}

\begin{document}
	\title{Impact of Non-Gaussian Beam Profiles in the Performance of Hadron Colliders}
	\author{S.~Papadopoulou}
	\affiliation{European Organization for Nuclear Research (CERN), CH-1211 Geneva 23, Switzerland}
	\affiliation{Department of Physics, University of Crete, P.O. Box 2208, GR-71003 Heraklion, Greece}
	\affiliation{Institute of Theoretical and Computational Physics (ITCP), GR-71003 Heraklion, Greece;}
	\author{F.~Antoniou}
	\affiliation{European Organization for Nuclear Research (CERN), CH-1211 Geneva 23, Switzerland}
	\author{T.~Argyropoulos}
	\affiliation{European Organization for Nuclear Research (CERN), CH-1211 Geneva 23, Switzerland}
	\author{M.~Hostettler}
	\affiliation{European Organization for Nuclear Research (CERN), CH-1211 Geneva 23, Switzerland}
	\author{Y.~Papaphilippou}
	\affiliation{European Organization for Nuclear Research (CERN), CH-1211 Geneva 23, Switzerland}
	\author{G.~Trad}
	\affiliation{European Organization for Nuclear Research (CERN), CH-1211 Geneva 23, Switzerland}
	
	\date{\today}
	
	\begin{abstract}

At the Large Hadron Collider (LHC), the interplay between a series of effects, including  intrabeam scattering (IBS), synchrotron radiation, longitudinal beam manipulations, two beam effects (beam-beam, e-cloud) and machine non-linearities, can change the population of the core and tails and lead to non-Gaussian beam distributions, at different periods during the beam cycle. By employing generalised distribution functions, it can be demonstrated that the modified non-Gaussian beam profiles have an impact in the beam emittance evolution by itself and thereby to the collider luminosity. This paper focuses on the estimation of beam distribution modification and the resulting rms beam size due to the combined effect of IBS and synchrotron radiation by employing a Monte-Carlo simulation code which is able to track 3D generalised particle distributions (SIRE). The code is first benchmarked over classical semi-analytical IBS theories and then compared with measurements from the LHC at injection and collision energies, including projections for the High-Luminosity LHC (HL-LHC) high brightness regime.  The impact of the distribution shape on the evolution of the bunch characteristics and machine performance is finally addressed. 
	\end{abstract}
	
	\maketitle

	\section{Introduction}
	\par The performance of a high-energy hadron collider such as the LHC is heavily based on the preservation of the injected emittances, under the influence of several degrading mechanisms. In this respect, an emittance evolution model was constructed including the effects of intrabeam scattering (IBS), synchrotron radiation, elastic scattering and luminosity burn-off (while at collision)~\cite{ref:model}. This simple model is based on semi-analytical approaches which assume Gaussian beam distributions, in particular for IBS. The bunch characteristics evolution predicted by this model revealed discrepancies, as compared to the measurements, translated to differences in the luminosity predicted by the model as compared to the experimental estimations~\cite{ref:model,ref:stefIPAC19,ref:upmodel}. One of the possible reasons for these differences could be attributed to the fact that the bunch profiles appear to be non-Gaussian  both at injection and collision energies, i.e. 450~GeV and 6.5~TeV respectively. The aim of this study is to quantify the impact of the beam distribution shape on the emittance and luminosity evolution of hadron colliders. In order to illustrate this, and employing generalised distribution functions, the luminosity of non-Gaussian beams is determined in a closed form. The generalisation of the luminosity estimate for arbitrary distributions does not only permit its comparison to the usual Gaussian beam estimate  but also the extension of classical results for the impact of non-Gaussian beam distributions to the luminosity (see Section~\ref{lumi}). This motivates the investigation of the emittance evolution beyond the classical analytical formulas for modelling IBS, which are based on 3D Gaussian beam assumptions~\cite{ref:BM}. In this respect, a Monte Carlo multi-particle simulation code for IBS and Radiation Effects (SIRE)~\cite{ref:SIRE, ref:IBSmartini2016}, is employed and compared to LHC data from Run2. A brief description of the code, and its benchmarking with the existing IBS analytical approaches and simulations for lepton rings is presented in Section~\ref{SIREgeneral}. Extending previous benchmarking studies for the LHC with respect to IBS theories~\cite{ref:vivlhc}, a detailed comparison of the Bjorken-Mtingwa (B-M) IBS theoretical model with the SIRE code for both injection and collision energies is presented for the nominal LHC using the Batch Compression Merging and Splitting (BCMS)~\cite{ref:bcms1,ref:bcms2} beam and the high luminosity LHC (HL-LHC)~\cite{ref:hllhc} beam parameters (Section~\ref{simulationsLHC}).  Finally, measured data from the LHC corresponding to non-Gaussian longitudinal beam profiles are compared with the expectations of the rms emittance evolution given by the theoretical B-M analytical formalism~\cite{ref:BM} and the SIRE code (Section~\ref{measured}). 
	
\section{Motivation- Impact of non-Gaussian distributions on Luminosity}
\label{lumi}
\par The performance of a collider is determined by the luminosity which, for two beams colliding head-on, is given by~\cite{ref:lumiconcept}:
\begin{equation}
\label{eq:lumi_int}
{\cal {L}} = 2 N_1 N_2 N_b f_{rev} {\int\limits}{\int\limits}{\int\limits}{\int_{-\infty}}^{\infty}
\rho_{1x}(x) \rho_{1y}(y) \rho_{1s}(s-s_0)  \rho_{2x}(x) \rho_{2y}(y) \rho_{2s}(s+s_0)~dx dy ds ds_0\:,
\end{equation}
with $N_{1,2}$ representing the number of particles per bunch for each beam, $N_b$ the total number of colliding bunches, $f_{rev}$ the revolution frequency and $\rho$ the beam density distribution functions for each plane and beam. 
\par Based on Eq.~\eqref{eq:lumi_int}, assuming Gaussian beams that collide head-on, the luminosity is expressed as~\cite{ref:lumiconcept}:
\begin{equation}
\label{eq:lumi_gauss2}
{\cal {L}}^G = \frac{ N_1N_2N_{b}f_{rev}}{4 \pi \sigma_x^G \sigma_y^G}\:.
\end{equation}
In order to achieve high luminosity, high intensity bunches and small beam sizes are required. The horizontal and vertical beam sizes of two colliding Gaussian bunches are given by:
\begin{equation}
\label{eq:beamsizes}
\sigma_x^G = \sqrt{\sigma_{1x}^2+\sigma_{2x}^2}~~~\mathrm{and}~~~\sigma_y^G = \sqrt{\sigma_{1y}^2+\sigma_{2y}^2}\:,
\end{equation}
where ($\sigma_{1x}$, $\sigma_{1y}$) and ($\sigma_{2x}$, $\sigma_{2y}$) are the transverse rms beam sizes of beam 1 and beam 2, respectively.
\par Based on the transverse and longitudinal bunch profile measurements, it has been observed that the particle distributions in the LHC, both at collision and injection energies, appear to have shapes that differ from the ones of a normal distribution~\cite{ref:miriam,ref:LHCsire,ref:Helga}. At the LHC injection energy, the emittance evolution is dominated by the IBS effect, both in the horizontal and in the longitudinal plane, while no effect is expected in the vertical plane~\cite{ref:Maria} where dispersion is minor. From Run~2 data, it is observed that in many cases the transverse bunch profiles appear to be non-Gaussian during the whole injection plateau~\cite{ref:miriam}. At the LHC collision energy, the IBS effect is weaker, while synchrotron radiation damping becomes more pronounced. The bunch profiles at collisions appear to have non-Gaussian tails, as well. In fact, during the energy ramp, the bunches that are blown up longitudinally in order to avoid instabilities due to the loss of Landau damping~\cite{ref:Baudrenghien}, arrive at the start of collisions with a clearly non-Gaussian shape~\cite{ref:Helga}. 

\par  By assuming that a particle distribution is Gaussian when this is not the case, not only the rms beam size may be underestimated or overestimated, but also its impact on performance parameters, such as the luminosity.  Therefore, it is important to use appropriate fitting functions (or some type of interpolation algorithm) on the beam profile in order to properly address this discrepancy. A generalized Gaussian function, called the $q$-Gaussian~\cite{ref:tsallis}, can be employed for fitting more accurately bunch profiles with shapes that differ from the ones of a normal distribution (see Appendix~\ref{qGaussian} for the properties of this distribution function). The parameter $q$ describes the weight of the tails as compared to the core, ranging from light tailed ones for $q<1$ (including the square distribution for $q \rightarrow-\infty$) and extending to a heavy tailed ones for $q>1$, passing through the Gaussian distributions in the limit of $q\rightarrow1$. This distribution is actually a stationary solution of a generalised Fokker-Plank equation which can cover a full spectrum of statistical behaviors of dynamical systems, from sub to super-diffusion Levy-type processes~\cite{ref:booktsallis}.
\begin{figure}[h]
		\includegraphics[clip, trim=0cm 0cm 0cm 0cm,width=0.35\textwidth]{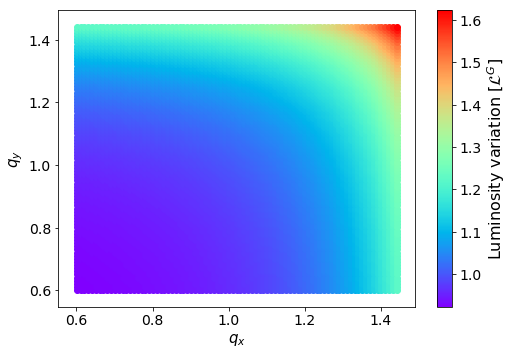}
		\captionsetup{justification=raggedright,singlelinecheck=false}
		\caption{Parameterization of the luminosity variation, normalised to the corresponding Gaussian luminosity value $\cal{L}$$^G$, with the weight of the transverse distribution tails given by the parameters $q_{x,y}$, for a constant q-Gaussian rms beam size.}
		\label{fig:lumi_qG}	
\end{figure}
\par In view of quantifying the impact of non-Gaussian distributions, the luminosity is estimated through Eq.~\eqref{eq:lumi_int} by using the specified probability density functions. Assuming that the two beams are identical and that they collide head-on, the luminosity  for $q$-Gaussian distribution functions in the transverse plane is given by:
\begin{equation}
\label{eq:lumi_qG_trans}
{\cal {L}} ^{qG} =\frac{ N_1N_2N_{b}f_{rev}}{4 \pi \sigma_x^{qG} \sigma_y^{qG}}
{\cal{I}}_x^{qG} {\cal {I}}_y^{qG}\:,
\end{equation}
for $\sigma_{x,y}^{qG}$ being the rms beam sizes (see Appendix~\ref{qGaussian}) in the transverse plane, for both beams. The factors ${\cal {I}}_{x,y}^{qG}$ which depend on the parameter $q$ in the respective planes and the details of the calculation are presented in Appendix~\ref{lumi_qGaussian}, together with the validation of the luminosity estimation for $q$-Gaussian distributions (shown in Fig.~\ref{fig:lumi_qconst}). 
By comparing this equation to the standard luminosity formula for Gaussian beams with identical rms sizes, the significance of the tail population contribution on luminosity can be established and parameterised through $q$. 
\begin{figure}[h]
	\begin{center}
		\includegraphics[width=0.3\textwidth]{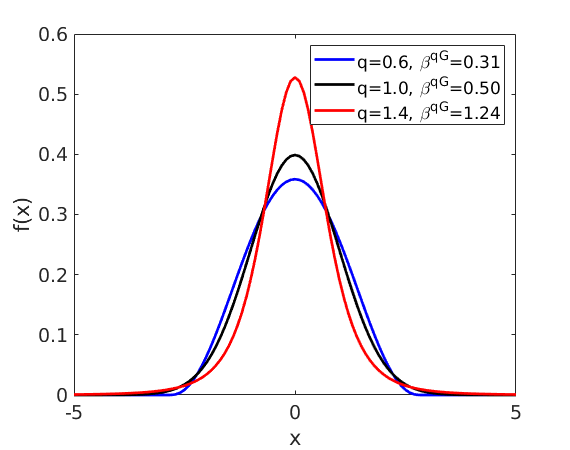}
		\captionsetup{justification=raggedright,singlelinecheck=false}
		\caption{The $q$-Gaussian density distribution function for a light tailed (blue, $q<1$), a Gaussian (black, $q=1$) and a heavy tailed (red, $q>1$) bunch profile, having identical $q$-Gaussian rms beam sizes.}
		\label{fig:distr_qG}	
	\end{center}
\end{figure}
This is illustrated in Figure~\ref{fig:lumi_qG} where the luminosity variation normalised to the corresponding one for Gaussian beams ($\cal{L}$$^G$) is parameterized with the parameter $q$ of the $q$-Gaussian fitting function, characterising the weight of the tails in the transverse plane, for fixed $q$-Gaussian rms beam sizes in all planes, assuming head on collisions (i.e. no dependence of the luminosity on the longitudinal beam size, see  Appendix~\ref{lumi_qGaussian}).
The bunch profiles corresponding to a light tailed ($q<1$), a Gaussian ($q=1$) and a heavy tailed ($q>1$) distribution, having identical beam sizes, are plotted in Fig.~\ref{fig:distr_qG}. As $q$ (and $\beta^{qG}$) vary for fixed $q$-Gaussian rms beam sizes (based on Eq.~\eqref{eq:sigma_qGauss} in Appendix~\ref{qGaussian}), the luminosity varies as well with respect to the one estimated for purely Gaussian beams. Practically, if the tails of a distribution differ by $10\%$ compared to the ones of a Gaussian distribution (i.e. $q=0.9$ or $q=1.1$), the luminosity value can be overestimated or underestimated by $5\%$. It is then clear that, for two beams colliding head-on, the shape of the transverse distributions has a significant impact to the estimated luminosity, in particular for the LHC experiments which target a $\sim2~\%$ accuracy in their estimates~\cite{ref:ilias}.
\begin{figure}[h]
	\begin{center}
		\includegraphics[width=0.35\textwidth]{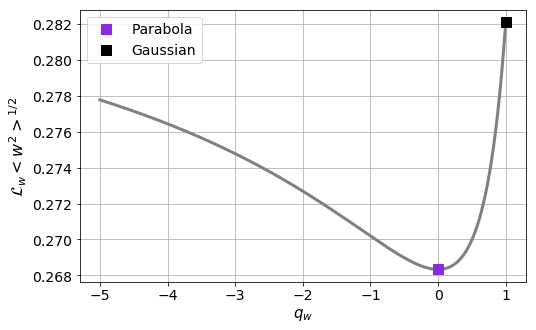}~~~
		\includegraphics[width=0.35\textwidth]{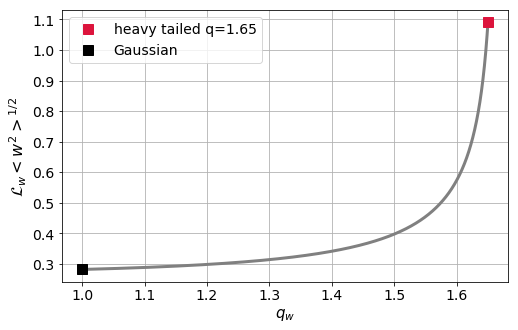}
		\captionsetup{justification=raggedright,singlelinecheck=false}
		\caption{In order to estimate the divergence of the luminosity for non-Gaussian distributions from the one for Gaussian densities, the ${\cal {L}}_w <w^2>^{1/2}$ (see Eq.~\ref{eq:lumiconstant1}) is plotted versus the weight of the tails $q$, for light tailed  (left) and the heavy tailed (right) distributions, i.e. $q<1$ and $q>1$, respectively. The points corresponding to a parabola and a Gaussian distribution, firstly presented in~\cite{ref:hereward, ref:potter}, are plotted.}
		\label{fig:int_q}	
	\end{center}
\end{figure}
The impact of non-Gaussian distributions on the luminosity was firstly discussed by  Hereward~\cite{ref:hereward}. In particular, the luminosity integrals were calculated for several distributions as the rectangular and the parabolic  which correspond to a $q$-Gaussian with $q\rightarrow-\infty$ and $q=0$, respectively. Assuming that the two beams are identical and that they collide head-on, these distributions were used as examples to estimate the discrepancy from the luminosity for Gaussian densities. This discrepancy is calculated after solving the ${\cal {L}}_w$ that is the integral over the density functions in one plane, being identical for both beams, as:
\begin{equation}
{\cal {L}}_w <w^2>^{1/2} = \sigma_{w} \int^\infty_{-\infty} \rho(w)^2dw\:,
\label{eq:lumiconstant1}
\end{equation}	
with $w=x,y$ and for $\sigma_{w}$ being the transverse rms beam size, since these solutions correspond to the transverse plane. It was found that~\cite{ref:hereward, ref:potter}:
\begin{equation}
{\cal {L}}_w <w^2>^{1/2} =
\begin{array}{cc}
\Bigg\{ & 
\begin{array}{l}
0.2887\:,~\mathrm{for~a~ rectangular~ distribution}\\
0.2683\:,~\mathrm{for~a~ parabolic~ distribution}~\\
\frac{1}{2\sqrt{\pi}} = 0.2821\:,~\mathrm{for~ a~ Gaussian~ distribution}~
\end{array}
\end{array}\:
\label{eq:lumiconstant}
\end{equation}	
In fact, this approach already identified the existence of a minimum for a light tailed parabolic distribution, which becomes obvious by employing the $q$-Gaussian, as observed in Fig.~\ref{fig:int_q}, where ${\cal {L}}_w <w^2>^{1/2}$ is plotted versus $q$  for $q<1$, i.e. light tails (left), and $q>1$, i.e. heavy tails (right). 
The results for $q$-Gaussian distributions (grey curves) are in perfect agreement with the case studies discussed in~\cite{ref:hereward}. This is also true for a rectangular distribution which corresponds to a $q$-Gaussian with $q\rightarrow-\infty$ and is beyond the range of the left plot of Fig.~\ref{fig:int_q}. For heavy tailed distributions, there is no upper limit for the constant of Eq.~\eqref{eq:lumiconstant}, as already inferred by Hereward~\cite{ref:hereward}. In Fig.~\ref{fig:int_q} (right) the case of a heavy tailed distribution with $q=1.65$ is denoted by a red square. 
The extreme case of $q\rightarrow5/3$  corresponds to a $q$-Gaussian whose rms size goes to infinity (i.e. Levy distributions, see Appendix~\ref{qGaussian}). 

\par The sensitivity of the luminosity on the distribution as generalised by employing the $q$-Gaussian function justifies the need of carefully studying the evolution of distributions in hadron colliders. For the LHC, a luminosity model was constructed~\cite{ref:model}. The evolution of the emittance includes the effects of IBS, synchrotron radiation, elastic scattering, betatron coupling, noise and burn-off. Although this model has a relative agreement with respect to the measured luminosities by the experiments (i.e. ATLAS~\cite{ref:Higgs} and CMS~\cite{ref:cms}), there is still some room for improvement~\cite{ref:upmodel}. Indeed, the emittance evolution for IBS was based on the module of MAD-X~\cite{ref:ibsmadx} following the B-M theory which assumes Gaussian beam distributions. The extension of this to non-Gaussian distributions as observed in the LHC may shed light to the origin of the remaining discrepancy between the model and the measurements. 
	
\section{Intrabeam scattering theories, observations and simulations}
\label{SIREgeneral}
\par One of the statistical processes causing a spreading of particles in phase space or a continuous increase of beam emittance is the small angle multiple Coulomb scattering, called Intrabeam scattering (IBS), which plays an important role in $e^+/e^-$ damping rings, high intensity/low energy light sources~\cite{ref:IBSls} and high intensity hadron~\cite{ref:IBShadron} and ion~\cite{ref:IBSion} circular machines.
The IBS theory for accelerators was firstly introduced by Piwinski~\cite{ref:Piwinskistand} and extended by Martini~\cite{ref:Martini}, establishing a formulation called 
the standard Piwinski method. Later, Bjorken and 
Mtingwa (BM)~\cite{ref:BM} used a different approach to describe the effect, taking into account the strong focusing effect. The Modified Piwinski method~\cite{ref:Bane} that includes the strong focusing effect, was developed by  Bane. Some approximations of these theories are the high energy one by Bane~\cite{ref:Bane} and the completely integrated modified 
Piwinski~\cite{ref:CIMP}.  
A different approach developed by Lebedev for hadron beams is based on a Boltzmann type integro-differential equation and includes  
betatron coupling~\cite{ref:Lebedev}. 	
\par The analytical models that describe the IBS  effect~\cite{ref:BM, ref:piwi} assume Gaussian beam distributions. 
The stationary solution of the Fokker-Planck equation is a particle distribution that is Gaussian in the phase space. However, taking into account the effects of IBS, radiation damping and quantum excitation but also other diffusive mechanisms~\cite{ref:Jowett}, there is no evidence that the distribution remains Gaussian. Therefore, it is important to  develop analytical formulas and simulation tools that calculate the interplay between these effects for any distribution. 
\par  For the Relativistic Heavy Ion Collider (RHIC), the IBS growth rates were calculated and benchmarked with experimental data using the distribution function evolution (based on the Fokker-Planck equations), extending the usual approach of employing a conventional Gaussian-like distribution~\cite{ref:wei}. In this respect, IBS growth rates were calculated for a bi-gaussian distribution, which was interesting for studying the possibility of using electron cooling in RHIC~\cite{ref:parzen}. Later, a model which is suitable for IBS calculations for arbitrary distribution functions and its comparison to experimental data was presented in~\cite{ref:fedotov}. The IBS effect was also studied for high-brightness electron linac beams which appear to be non-Gaussian, especially in the longitudinal plane~\cite{ref:xiao}.
For low-emittance high-intensity electron storage rings, the interplay between intrabeam scattering and wake-field forces is discussed in~\cite{ref:venturini}. In particular, the calculation of the IBS growth rates and the estimation of the emittance growth is discussed in detail in~\cite{ref:lebedev2}, where the importance of knowing the formation of the distribution tails is underlined, referring also to the use of Monte-Carlo methods.
\par In order to simulate the impact of a distribution shape on the emittance evolution 
when considering IBS and radiation effects the SIRE (Software for IBS and Radiation Effects)~\cite{ref:SIRE,ref:IBSmartini2016} code was developed by Vivoli and Martini 
at CERN. A similar Monte Carlo approach (IBStrack~\cite{ref:Theo}) was implemented also in the collective effects  simulation tool CMAD~\cite{ref:CMAD, ref:CMADibs}. Both algorithms have as their basis  MOCAC (MOnte CArlo Code), a Monte-Carlo code developed by Zenkevich and collaborators~\cite{ref:MOCAC1, ref:MOCAC2}, which calculates the IBS effect for arbitrary distributions, by representing the beam through a large number of macro-particles occupying points in the 6-dimensional phase space.  Being an extension of MOCAC, SIRE was developed to simulate the evolution of the beam particle distributions, taking into account the effects of IBS, synchrotron radiation and quantum excitation. For the evolution of the LHC bunch parameters and the shape of the bunch profiles presented in this paper, the simulations are performed using SIRE. As the physics and implementation details of the code are extensively explained in~\cite{ref:IBSmartini2016}, we only briefly summaries some of its key features.

\subsection{Software for IBS and Radiations Effects (SIRE)}
\par For the IBS calculations, SIRE~\cite{ref:SIRE,ref:IBSmartini2016} uses the classical Rutherford cross section, which is closer to the Piwinski formalism~\cite{ref:piwi}. It uses as  input the optics functions at different locations of the lattice in order to determine the trajectories of the particles in phase space. Instead of using the 6 coordinates for position and momentum, the two Courant-Snyder and longitudinal invariants and the 3 phases (betatron and synchrotron) are used. For a linear ring, the 3 invariants are conserved between points around the lattice and can only be changed by the effects of IBS, synchrotron radiation and quantum excitation, while the phases are chosen randomly at each given point of the lattice. The time steps for which the IBS and radiation effects are called should be specified such that they are larger than the revolution time and smaller than the damping/growth times. Dividing the total time by the time steps shows how frequently the IBS, synchrotron radiation and quantum excitation routines are called. The quantum excitation is implemented by adding to the 6 coordinates of each macro-particle a Gaussian random noise component.
\par Depending on the elapsed time, the synchrotron radiation damping acts on the invariants of the macro-particles as an exponential decrement.  The routine introduced for this reason is called after the calculation of the IBS effect at each iteration. Using small iteration time steps $dt$ (which are much smaller than the damping times and for which the emittances change adiabatically), the evolution of the transverse emittance and energy spread due to the effects of IBS and synchrotron radiation can be obtained by solving the coupled
differential equations:
\begin{equation}
\begin{aligned}
\dfrac{d\epsilon_{x,y}}{dt}&=\dfrac{-2(\epsilon_{x,y}-\epsilon_{{x,y}0})}{\tau_{x,y}}+\dfrac{2\epsilon_{x,y}}{T_{x,y}}\:,\\
\dfrac{d\sigma_{p}}{dt}&=\dfrac{-(\sigma_{p}-\sigma_{{p}0})}{\tau_{p}}+\dfrac{\sigma_{p}}{T_{p}}\:,
\label{eq:evol}
\end{aligned}
\end{equation}
with $\epsilon_{{x,y}0}$ and $\sigma_{{p}0}$ being the zero current (without the effect of IBS) equilibrium transverse emittances and energy spread, respectively. The $\tau_{x,y}$, $\tau_{p}$ are the synchrotron radiation damping times and $T_{x,y}$, $T_{p}$ the IBS growth times. 

\par The algorithm SIRE uses to calculate IBS is similar to that implemented in MOCAC, where the beam is represented by a large number of macro-particles occupying points in the 6-dimensional phase space. The default distribution defined in SIRE by using a random number generator, is the Gaussian and is given in action angle variables. 
In order to introduce a different distribution,  the proper random deviates should be generated or the action angle variables of all macroparticles for the desired distribution should be provided. After specifying the total beam population and the number of macro-particles, the initial distribution can be tracked. The particle distribution in all planes can be saved as often as requested during the simulation time. Currently, the output file gives the evolution of the emittance in all planes for the specified time steps.

\par After providing the beam distribution and the optics along a lattice, the beam is geometrically divided according to the specified number of cells for each plane. The macro-particles are assigned to each cell according to their geometrical position. For each lattice point defined in the optics file, the 3 phases of each macro-particle are randomly chosen and the position and momentum of the macro-particles are calculated.  Based on the classical Rutherford cross section, intra-beam collisions  between pairs of macro-particles are calculated in each cell. The momentum of particles is changed  due to scattering. The number of	macro-particles and cells, i.e. the number of collisions each macro-particle experiences, is chosen so as to give accurate results for a reasonable computational time. The scattering angles for each collision are determined. In order to get the mean value of the emittance and momentum deviation changes for all particles, we have to integrate over the phase space volume of betatron coordinates, momentum deviations and azimuthal positions of the interacting particles.
The beam distribution is then updated based on the new invariants of the macro-particles. For a specified number of time steps which practically shows how frequently the IBS and synchrotron radiation routines are called, the beam distribution is updated and the rms beam emittances are recomputed, giving finally as output the emittance evolution in time. The simulation proceeds to the next lattice point and continues until the end time is reached.

\par A lattice compression technique named ``lattice recurrences'' has been implemented to speed up the calculations~\cite{ref:SIRE}. Since the increase of the invariants due to IBS is linear to the first order in the traveling time along an element, the IBS kicks with optics functions differing less than a specified precision value are considered equivalent. For the corresponding group of elements, the IBS effect is evaluated only once, resulting to a reduction of the computational time.

\subsection{The logarithmic Coulomb factor}	
\par The IBS growth times have a complicated dependence on the beam properties, due to the coupling of the three planes through dispersion. 
Some of these properties are the bunch charge and energy, the beam optics and the equilibrium rms horizontal, vertical emittances and the energy spread. The IBS growth times depend also on a logarithmic Coulomb factor which is used to include the contribution of events having a very large and very small impact parameter. The typical way of computing a log factor overemphasises the importance of the very small impact parameter events, for which the tails of the steady-state bunch distributions are non-Gaussian. In the high energy approximation by Bane~\cite{ref:IBSls}, in order to describe the size of the core of the bunch, the Coulomb log factor is calculated as was first proposed by Raubenheimer~\cite{ref:Raubenheimer}, i.e. based on a boundary between the contribution to the core and the tails.  In B-M, Bane and CIMP methods, the Coulomb factor is defined as the ratio of the maximum $r_{\mathrm{max}}$ to the minimum $r_{\mathrm{min}}$ impact parameter in the collision of two particles in the bunch, that is $(log)\equiv \ln(r_{\mathrm{max}}/r_{\mathrm{min}})$. For typical  flat beams, the $r_{\mathrm{max}}$ is taken to be equal to the vertical beam size $\sigma_y$, while $r_{\mathrm{min}}$ is taken to be $r_{\mathrm{min}}=r_0\beta_x/(\gamma^2\epsilon_x)$, with $r_0$ being the classical particle radius. Then, the Coulomb factor can be written as:
\begin{equation}
\label{eq:coullog}
(log)=\mathrm{ln}\left( \frac{\gamma^2 \epsilon_x \sqrt{\beta_y \epsilon_y}}{r_0 \beta_x} \right)\:.
\end{equation}
The formalism by Piwinski always seems to underestimate the IBS effect with respect to the other theoretical models. What diversifies Piwinski's method, is the different definition of the Coulomb factor. In that approach, the maximum impact parameter which is typically taken as the vertical beam size appears. In the high energy limit, with $d$ being the maximum impact parameter, the Coulomb $(log)$ for Piwinski can be written as~\cite{ref:Banecmp}:
\begin{equation}
\label{eq:coullogP}
(log)=\mathrm{ln}\left( \frac{d \sigma_x^2}{4 r_0 \alpha^2} \right)\:,
\end{equation}
where $a=\frac{\sigma_x}{\gamma}\sqrt{\frac{\beta_x}{\epsilon_x}}$.
Comparing the $(log)$ factors of Eq.~\eqref{eq:coullog} and~\eqref{eq:coullogP}, we find that condition $d=4 \sigma_y$ in order for the Piwinski approach agreeing with the other models. SIRE uses the ``binary collision map" algorithm, conceived by Zenkevich, which allows to reduce the effects of the continuous time dynamical IBS system to a discrete time map in momentum space. For the binary collision events, the maximum impact parameter is taken as the beam height. 


%
	\subsection{Benchmarking of IBS theoretical model with SIRE}
	\par The performance of hadron machines is limited by the IBS effect causing emittance growth. For lepton machines such as future linear collider Damping Rings, new generation light sources and B-factories, the IBS effect can also be predominant. It is thus important to study the IBS theories in the presence of synchrotron radiation and quantum excitation and benchmark the existing theoretical models and tracking codes with experimental data. In this way, the codes limitations can be identified so that to apply the necessary improvements in order to get better predictions for a machine's operation.  

	\par
\begin{figure}[h]
			\centering
			\includegraphics[width=0.28\textwidth]{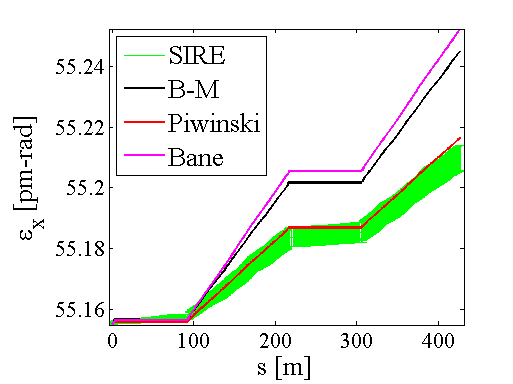}
			\includegraphics[width=0.28\textwidth]{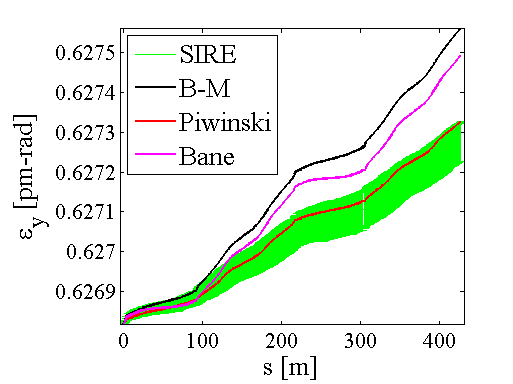}
			\includegraphics[width=0.28\textwidth]{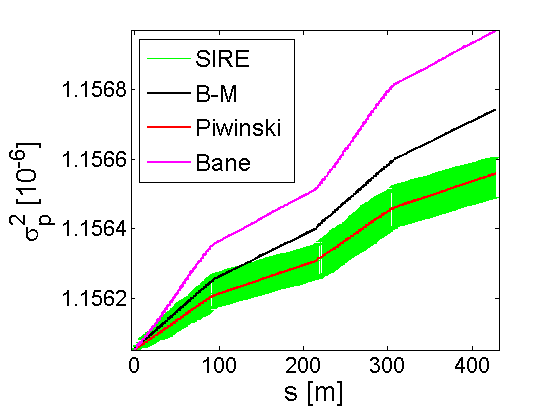}
			\captionsetup{justification=raggedright,singlelinecheck=false}
			\caption{One turn comparison for the horizontal (left) and vertical (middle) emittance and
			energy spread (right) between the tracking code SIRE and the theoretical models Bjorken-Mtingwa (B-M), Piwinski and
			Bane for the CLIC DR lattice.}
			\label{fig:clic}
		\end{figure}

 A benchmarking of the IBS theoretical models with Monte-Carlo codes is presented in~\cite{ref:Fanthesis} for lepton rings. The comparison between different theoretical models and SIRE is discussed for the Compact Linear Collider (CLIC) damping ring (DR)~\cite{ref:clic}, having ultra-low emittances which are strongly dominated by IBS, in the presence of synchrotron radiation and quantum excitation. Results of this comparison are presented in Fig.~\ref{fig:clic}, for one turn of the DR lattice. Due to the fact that in SIRE the generation of the distribution is based on a random number generator, the tracking simulations were performed several times, resulting in the one standard deviation error-bars (plotted in green). The classical
	formalism of Piwinski (red colored) and SIRE are in perfect agreement, as was expected, since SIRE uses the classical Rutherford cross section which is closer to the Piwinski formalism.  The Bjorken-Mtingwa (black colored) and Bane (magenta colored) formalisms overestimate the effect compared to Piwinski method mainly to a mismatch of the Coulomb factor used in the different approaches (see Eq.~\eqref{eq:coullogP}).
	   
	\par The IBS theoretical models have been studied in detail and benchmarked with experimental data also for hadron beams over the years~\cite{ref:IBShadron, ref:IBSion}. A comparison of the LHC data with simulations performed with SIRE is discussed in~\cite{ref:vivlhc, ref:LHCsire}. In this paper, the SIRE simulations, as well as the benchmarking with the B-M formalism and experimental data are discussed in detail for the LHC and the High Luminosity LHC (HL-LHC)~\cite{ref:hllhc}.
		
	\section{Simulations performed with SIRE for the LHC}
	\label{simulationsLHC}
	\par In order to understand the evolution of the bunch characteristics, based on the bunch profile observations, it is important to study the interplay between IBS and radiation effects (synchrotron radiation and quantum excitation) during the full LHC cycle. This is done using  SIRE for two cases which are important for the current and future machine performance; the nominal BCMS~\cite{ref:bcms1,ref:bcms2} and the HL-LHC~\cite{ref:hllhc} parameters. 
	Despite some blow-up in the LHC during the ramp, it is observed that the BCMS beam gives an increase in peak luminosity of around 20\%. The HL-LHC is the major LHC upgrade aiming to increase integrated luminosity by at least a factor of 10 compared to the nominal LHC design value (from 300 to 3000 fb$^{-1}$). In order to achieve that, the bunch population needs to be increased  and the transverse beam size at the collision points has to be lowered. 
	\par Apart from the IBS and synchrotron radiation
 which are the dominant effects for the emittance evolution in the LHC, a combination of other diffusion mechanisms, like the beam-beam effect, electron-cloud, noise (due to the power converters, the transverse damper, the crab cavities, etc.), and other non-linearities cause emittance growth and/or particle losses~\cite{ref:Lamont}. Despite the fact that these mechanisms are not included in SIRE, it is possible to add empirically (i.e. based on observations) their contribution. Practically, there is the option of adding or complementing the variation of the bunch parameters in time.
	In fact, the simulation studies presented in this paper for the LHC are focused on the 3$\sigma$ range of the particle distributions and therefore, mechanisms 
	which concern the far tail regime are not taken into account as they are more important for the lifetime of the beam then on distribution evolution
.
	\subsection{Reduced lattice}
		\par As mentioned earlier, one of the inputs required by SIRE are the optical functions along the ring. 	As the LHC is a very long accelerator of about 27~km, with a very large number of elements in the sequence (more than 11000), SIRE requires an extremely long computational time to track the distribution for all the elements along the ring.   Aiming to reduce the computational time, a study was first performed in order to identify the optimal minimum number of critical IBS kicks around the lattice, without affecting the overall effect. 
		The IBS growth rates were firstly calculated for the full optics of the LHC, using the IBS module of the Methodical Accelerator Design code (MAD-X)~\cite{ref:ibsmadx}
		which is based on the Bjorken-Mtingwa formalism. 
		\begin{figure}[h]
			\centering
			\includegraphics[width=0.6\textwidth]{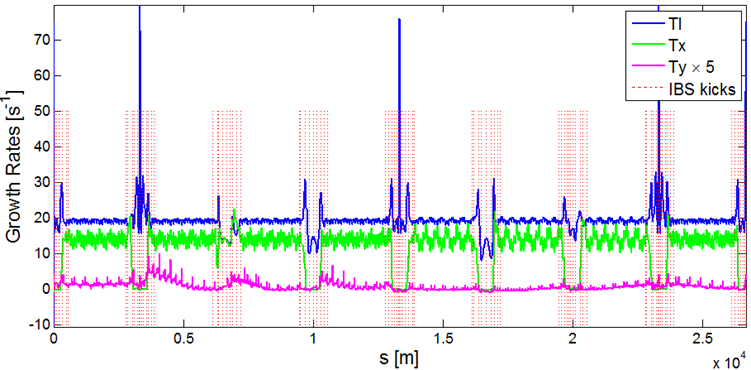}
			\captionsetup{justification=raggedright,singlelinecheck=false}
			\caption{The IBS growth rates along the LHC in all three planes: longitudinal (green), the horizontal (blue) and the vertical (magenta). The IBS kicks that are noted with red dashed lines, represent the positions of the 92 elements that compose the reduced lattice.}
			\label{fig:ibskicks}
		\end{figure}
		Figure~\ref{fig:ibskicks} shows the IBS growth rates in the longitudinal (green), the horizontal (blue) and the vertical (magenta) plane. Taking into account the strong IBS kicks along the ring, various lattices with a reduced number of elements, including the case of the smooth lattice approximation, were tested. Then, using the IBS module of MAD-X, the emittance evolution was calculated for several sets of beam parameters to assure that the choice of the elements is valid for a wide range of regimes, for which the IBS impact may be weaker or stronger.   Finally, the optimal lattice chosen consists of only 92 elements whose positions are denoted by red dashed lines in Fig.~\ref{fig:ibskicks}.  
		\begin{figure}[h]
			\centering
			\includegraphics[width=0.35\textwidth]{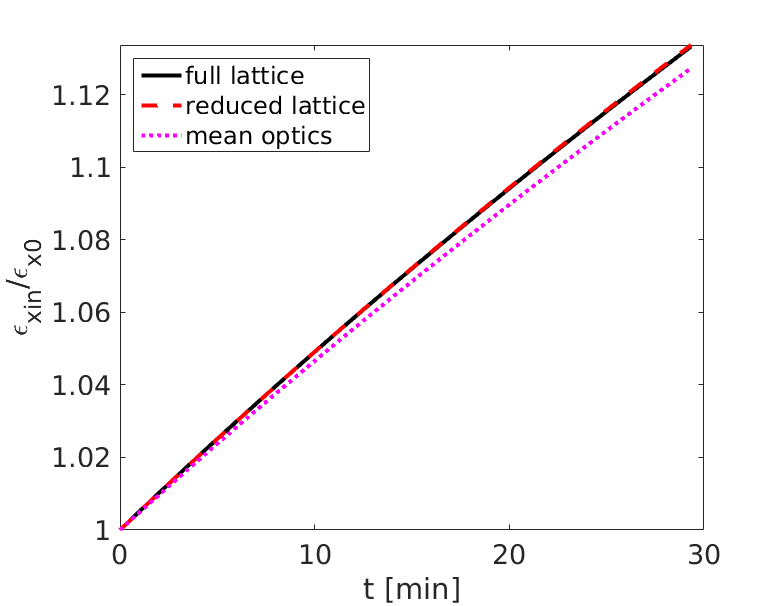}
			~~~~~
			\includegraphics[width=0.35\textwidth]{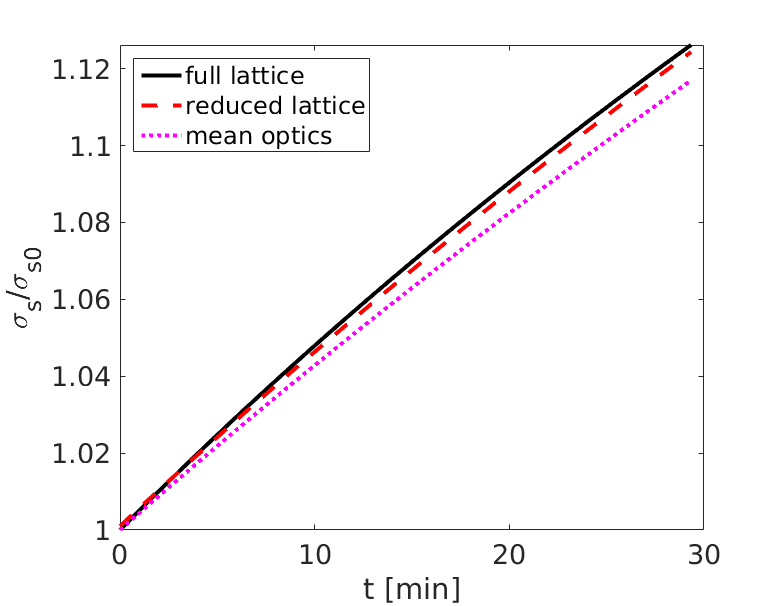}
			\captionsetup{justification=raggedright,singlelinecheck=false}
			\caption{The growth of the horizontal emittance (left) and bunch length (right) due to IBS, as computed by MAD-X  with the Bjorken-Mtingwa formalism, in a time period of 30 min at injection, when considering the full lattice (black solid line), the reduced lattice (red dashed line) and the  smooth lattice approximation- mean optics (magenta dotted line).}
			\label{fig:reduced}
		\end{figure}
		\par Figure~\ref{fig:reduced} shows the emittance (left) and the bunch length (right) growth during 30~min at  injection energy, for the nominal BCMS beams, with initial emittance and $4\sigma$ bunch length that are respectively $\epsilon_{x0}=1.5$~$\mu m$$rad$ and $\sigma_{s0}=1$~ns, having a bunch population of $1.2\times10^{11}$ protons. The black solid line refers to the case of the full lattice, while the red dashed one to the reduced lattice with the 92 elements. The magenta dotted line corresponds to the case of the smooth lattice approximation for which a lattice with a unique element, having the optics that represent in the best possible way the mean optics of the full lattice, is considered.   
		The agreement of the full and the reduced lattice is very good in all planes. On the other hand, by using the smooth lattice approximation the IBS effect is slightly underestimated, in particular, in the longitudinal plane. In the next, since the results for the reduced and the full lattice agree also in SIRE, the reduced lattice is used as an input for the simulation code. After choosing the optimal number of cells and macro-particles, the computational time in the case of the reduced lattice is almost 20 times shorter than the one of the full LHC lattice.
		
		
		\subsection{Convergence studies}
		\par For a specified set of input beam parameters, various scans should be performed for different combinations of number of macro-particles and cells in order to find the optimal values which provide a fast tracking and at the same time, guarantee that the scattering process leads to accurate results. In these terms, in order to avoid having a very small number of macro-particles per cell, the total number of cells is calculated based on the optimal minimum number of macro-particles per cell. For $n_x$, $n_y$, $n_z$ being the number of cells in the horizontal, vertical and longitudinal plane, respectively, it is assumed that in the transverse plane there is a correlation between the number of cells ratio and the beam sizes ratio, meaning that  $n_x/n_y$=$\sigma_x$/$\sigma_y$. Therefore, for $n_{mp}$ being the total number of macro-particles, the number of macro-particles per cell is:
		\begin{equation}
		n_{mp}/cell=\dfrac{n_{mp}}{n_x n_y n_z}=\dfrac{n_{mp}}{n_x^2 (\dfrac{\sigma_y}{\sigma_x}) n_z}\:.
		\label{eq:MPperCell}
		\end{equation}
		\par
\begin{figure}[h]
			\centering
			\includegraphics[width=0.3\textwidth]{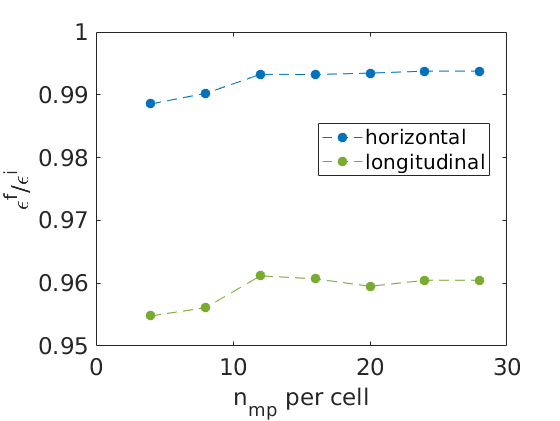}
			\includegraphics[width=0.3\textwidth]{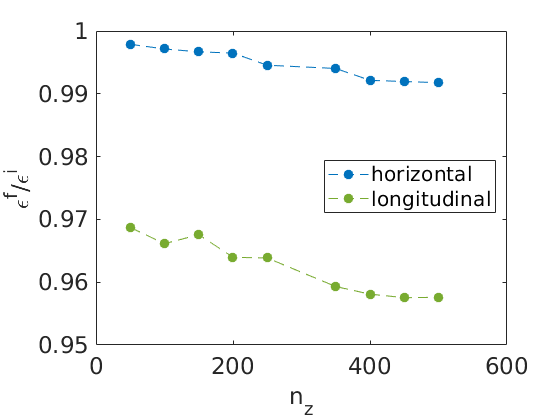}
			\includegraphics[width=0.3\textwidth]{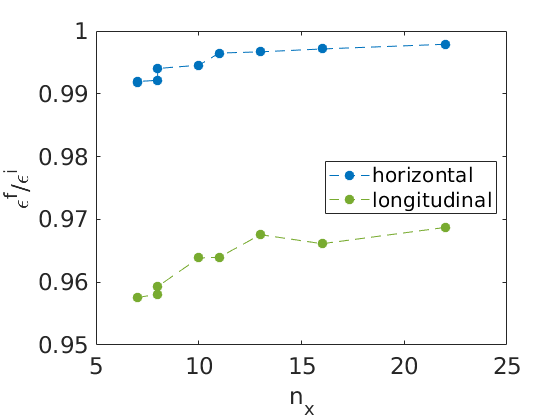}
			\captionsetup{justification=raggedright,singlelinecheck=false}
			\caption{\small {The dependence of the horizontal (blue) and longitudinal (green) emittance variation on the number of macro-particles per cell (left), on the number of cells in the longitudinal plane (middle) and in the horizontal plane (right), for a specific time period.}}
			\label{fig:scanMPperCell}
			\end{figure}

 A scanning of the total number of cells is performed for an example set of beam parameters to be used as an input for tracking. Based on Eq.~\eqref{eq:MPperCell}, by keeping the total number of macro-particles constant, the different combinations of cell numbers determines the number of macro-particles per cell.  Figure~\ref{fig:scanMPperCell} (left) shows the dependence of the emittance variation (ratio of final versus initial value) in the horizontal (blue) and longitudinal (green) plane on the number of macro-particles per cell, for a specified time duration. The value of the number of macro-particles per cell after which the variation of the emittances in both planes remains constant is chosen as the optimal minimum value.  After specifying this value, a scanning is performed  for a fixed number of macro-particles, in order to choose the number of cells to be used, firstly in the longitudinal plane.
					

		Then, the number of cells in the horizontal plane can be calculated using Eq.~\eqref{eq:MPperCell}, when knowing the ratio of the beam sizes in the transverse plane~\footnote{Here it is assumed that the ratio of the transverse beam sizes is initially $\sigma_x$/$\sigma_y=1$.}. In Fig.~\ref{fig:scanMPperCell}, the dependence of the emittances variation is plotted versus the number of cells in the longitudinal (middle) and horizontal (right) plane, for a specific time duration. It can be noticed that the variation of the emittances remains constant after a certain number of cells in the longitudinal and horizontal plane that is, for the example set of beam parameters, 350 and 13 cells, respectively. Finally, the number of cells in the vertical plane can be calculated by $n_y=n_x/(\sigma_x/\sigma_y)$.					
		
		\subsection{Benchmarking of SIRE with the B-M IBS theoretical model} 
		\par  SIRE has the advantage to accept any type of distribution as an input. If requested, it also gives as output the distribution at any stage of the tracking. In order to  benchmark the code with the analytical formulation of B-M for the LHC, a Gaussian distribution was tracked for two sets of bunch parameters which are summarized in Table~\ref{tab:params} for both the injection ($450$ GeV) and collision energy ($6.5$ TeV). The first case corresponds to the nominal BCMS~\cite{ref:bcms1,ref:bcms2} LHC beams, having a significantly lower transverse beam size with respect to the nominal production scheme. The second case corresponds to the HL-LHC~\cite{ref:hllhc, ref:hllhc2} parameters, for which the bunch population is very high. The input optics functions used for tracking correspond to the ones of the aforementioned reduced lattice.    
		
		\begin{table}[h]
			\caption{Nominal (BCMS) and HL-LHC parameters, at injection energy ($450$ GeV) and at collision energy ($6.5$ TeV).}
			\label{tab:params}
			\centering
			\begin{tabular}{lcccc}\hline\hline
			\multirow{2}{*}{\textbf{IBS  growths}}& \multicolumn{2}{c}{\textbf{Injection energy}}	& \multicolumn{2}{c}{\textbf{Collision energy}} 
				\\
				& LHC (BCMS)     	& HL-LHC 	& LHC (BCMS)     	& HL-LHC \\\hline
				$\epsilon_{x,y}$~[$\mu$m.rad]		& 1.5	& 2.0	& 2.5 	& 2.5		\\
				4$\sigma$ bunch length~[ns]			& 1.0	& 1.2	& 1.0 	& 1.2		\\
				Bunch population~[$10^{11}$]		& 1.2	& 2.3	& 1.1	& 2.2	\\\hline\hline
			\end{tabular}
		\end{table} 
		
		\subsubsection{At the LHC injection energy (450~GeV)} 
\par  The evolution of the horizontal emittance (left), the vertical emittance (middle) and energy spread (right) after 1~h at injection energy ($450$ GeV), 	 
		\begin{figure}[h]
			\centering
			\includegraphics[width=0.325\textwidth]{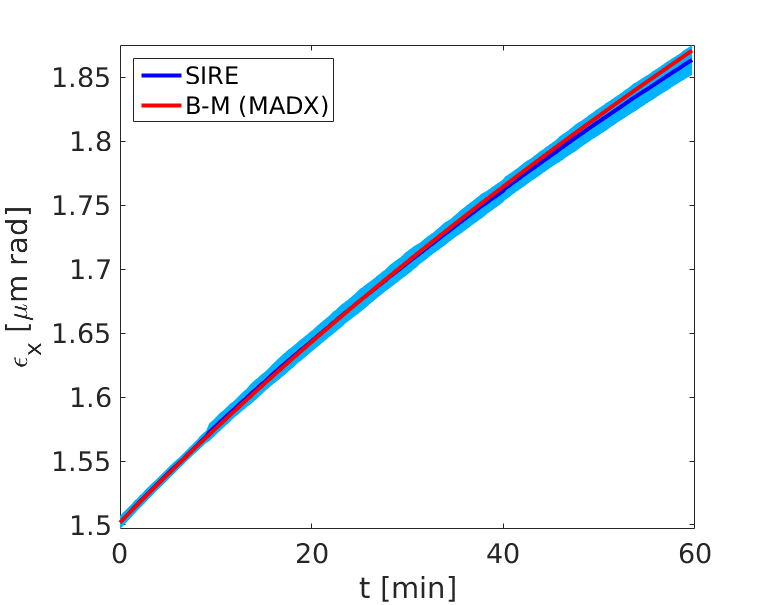}
			\includegraphics[width=0.325\textwidth]{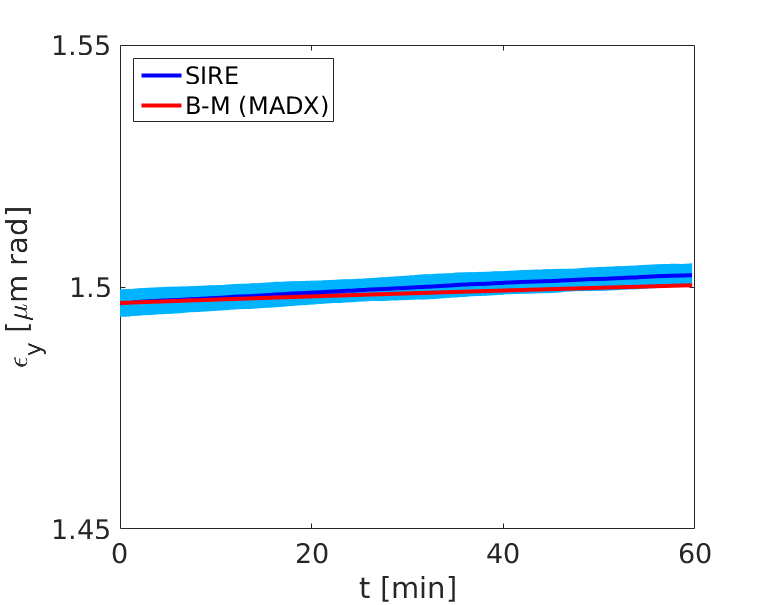}
			\includegraphics[width=0.325\textwidth]{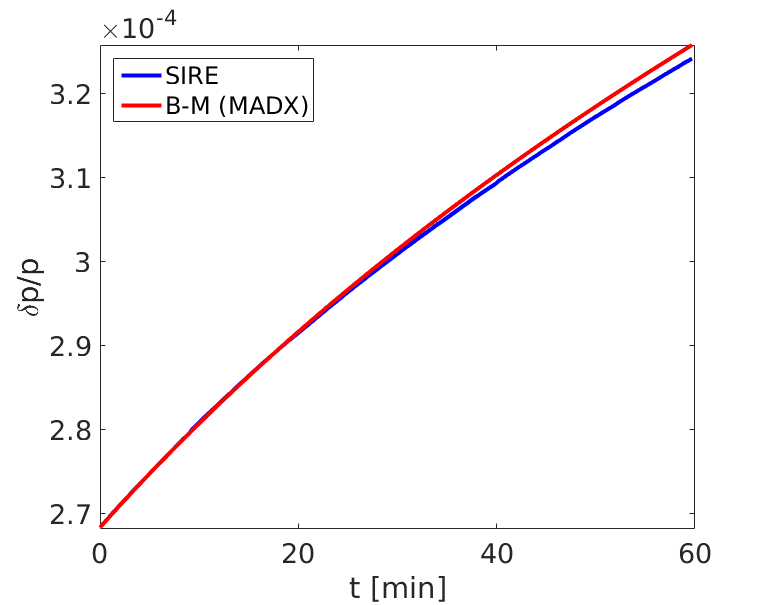}
			\captionsetup{justification=raggedright,singlelinecheck=false}
			\caption{The growth of the horizontal (left) and vertical (middle) emittance and energy spread (right) due to IBS, in a time period of 1~h at the injection energy of the LHC  ($450$ GeV) for the nominal parameters, as computed by the SIRE code (blue line) and the Bjorken-Mtingwa analytical formalism in MAD-X (red line).}
			\label{fig:sirebench_nom_fb}
		\end{figure}
		\begin{figure}[h]
			\centering
			\includegraphics[width=0.325\textwidth]{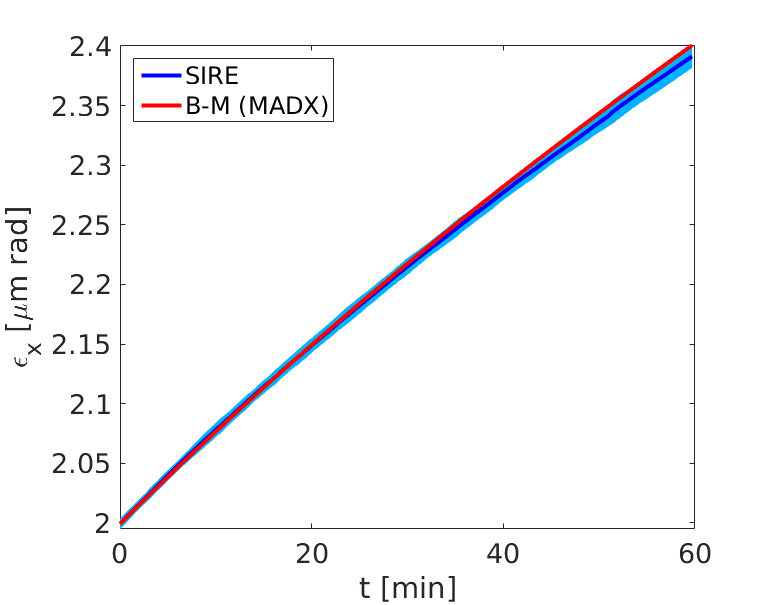}
			\includegraphics[width=0.325\textwidth]{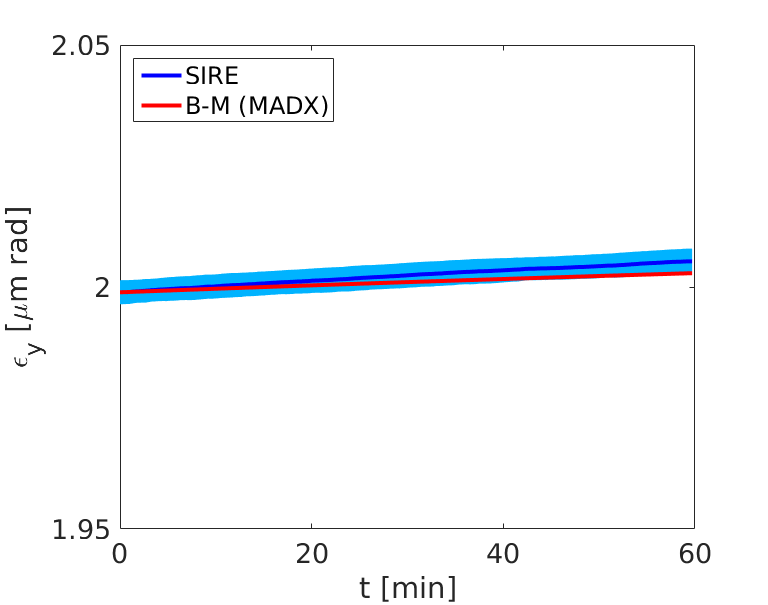}
			\includegraphics[width=0.325\textwidth]{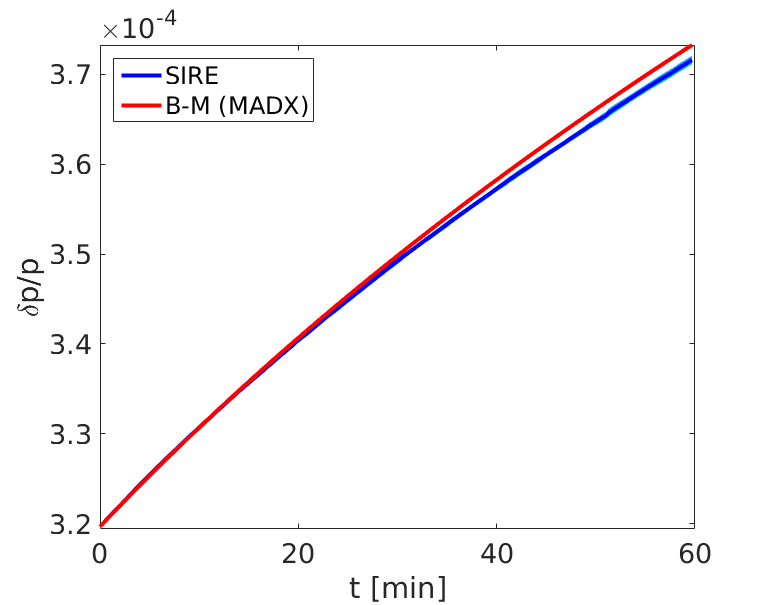}
			\captionsetup{justification=raggedright,singlelinecheck=false}
			\caption{The growth of the horizontal (left) and vertical (middle) emittance and energy spread (right) due to IBS, in a time period of 1~h at the injection energy of the LHC  ($450$ GeV) for the HL-LHC parameters, as computed by the SIRE (blue line) and the Bjorken-Mtingwa analytical formalism in MAD-X (red line).}
			\label{fig:sirebench_hilumi_fb}
		\end{figure}
where the IBS effect is dominant, are presented in Fig.~\ref{fig:sirebench_nom_fb} for the nominal BCMS case and in Fig.~\ref{fig:sirebench_hilumi_fb} for the HL-LHC parameters.		The red and the blue lines correspond to the analytical calculations of the MAD-X~\cite{ref:MADX} IBS routine (based on the B-M formalism) and to the SIRE results, respectively.  	
		\begin{table}[h]
			\caption{IBS  growths of the transverse emittances and energy spread during 1~h at injection energy ($450$ GeV).}
			\label{tab:IBSgrowths}
			\centering
			\begin{tabular}{lcccc}\hline\hline
				\multirow{2}{*}{\textbf{IBS  growths}}		& \multicolumn{2}{c}{\textbf{Nominal (BCMS)}}	& \multicolumn{2}{c}{\textbf{HL-LHC}} 
				\\
				&  $MAD-X$     	& $SIRE$ 	&  $MAD-X$     	& $SIRE$ \\\hline
				$d\epsilon_x/\epsilon_{x0}$ [\%]		& 24.6		& 24.1		& 20.1 	& 19.6		\\
				$d\epsilon_y/\epsilon_{y0}$ [\%]			& 0.2		& 0.4		& 0.2 	& 0.3		\\
				$d\sigma_l/\sigma_{l0}$	   [\%]		& 21.4	& 20.8	& 16.8	 & 16.2	\\\hline\hline
			\end{tabular}
		\end{table} 					
		Due to the fact that in SIRE the generation of the distribution is based on a random number generator, the tracking simulations were performed several times, resulting in the two standard deviation spread that are plotted in light blue. Table~\ref{tab:IBSgrowths} summarizes the IBS growth of the transverse emittances and energy spread,  for the nominal BCMS and HL-LHC parameters, as computed by the SIRE code and the B-M analytical formalism in MAD-X.
		\begin{figure}[h!]
			\centering
			\includegraphics[width=0.325\textwidth]{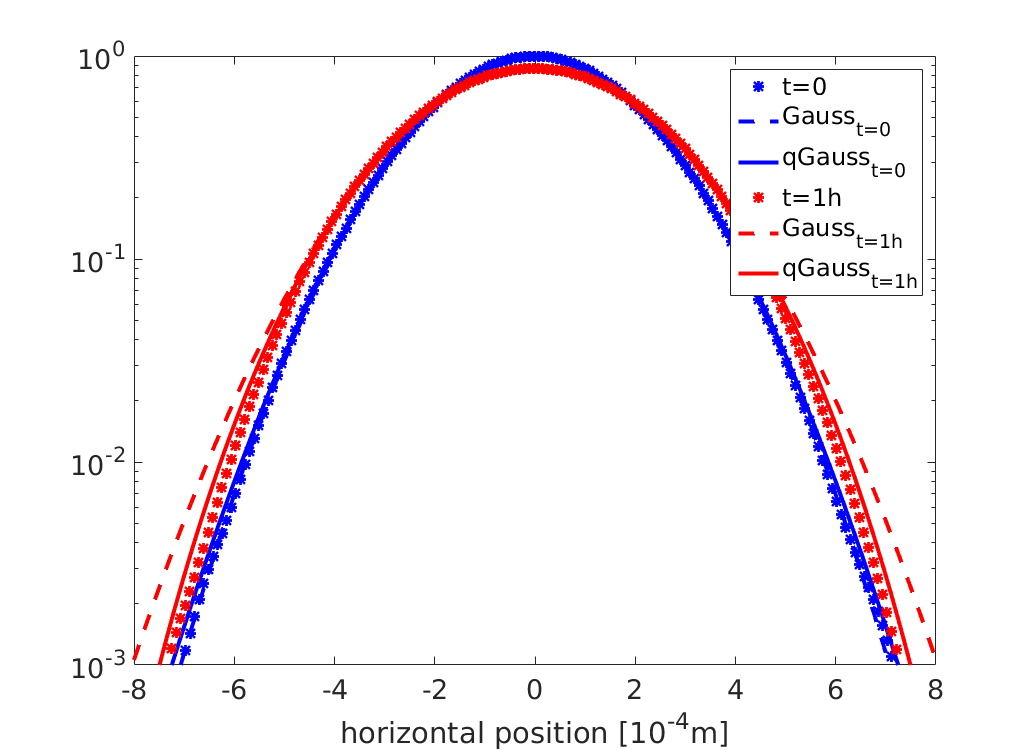}
			\includegraphics[width=0.325\textwidth]{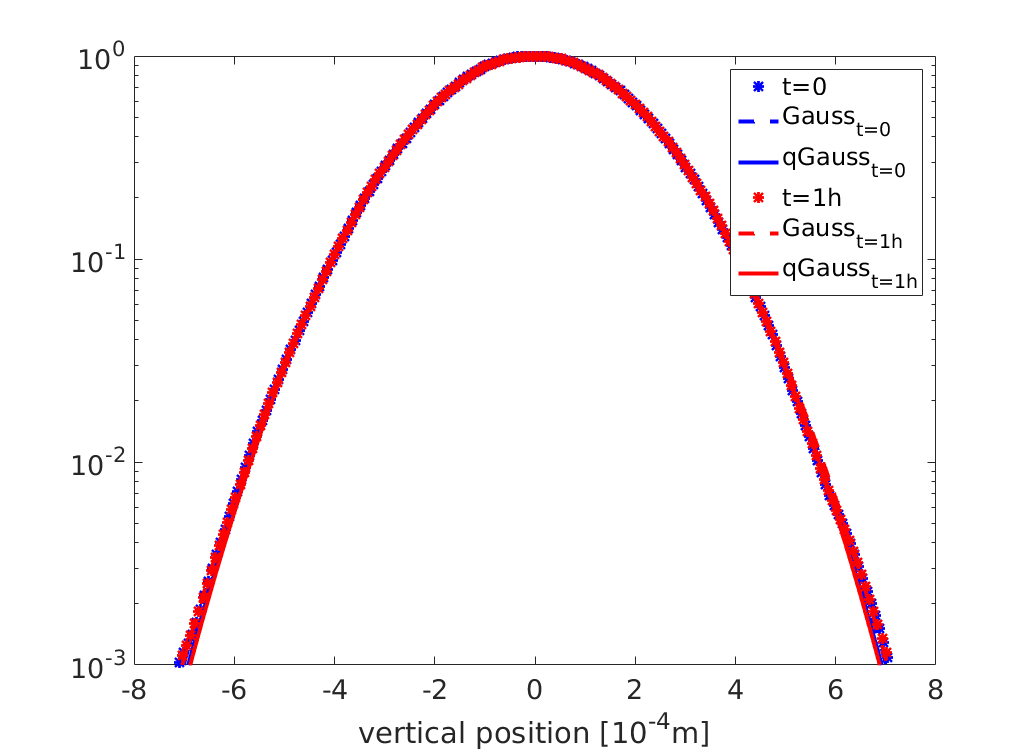}
			\includegraphics[width=0.325\textwidth]{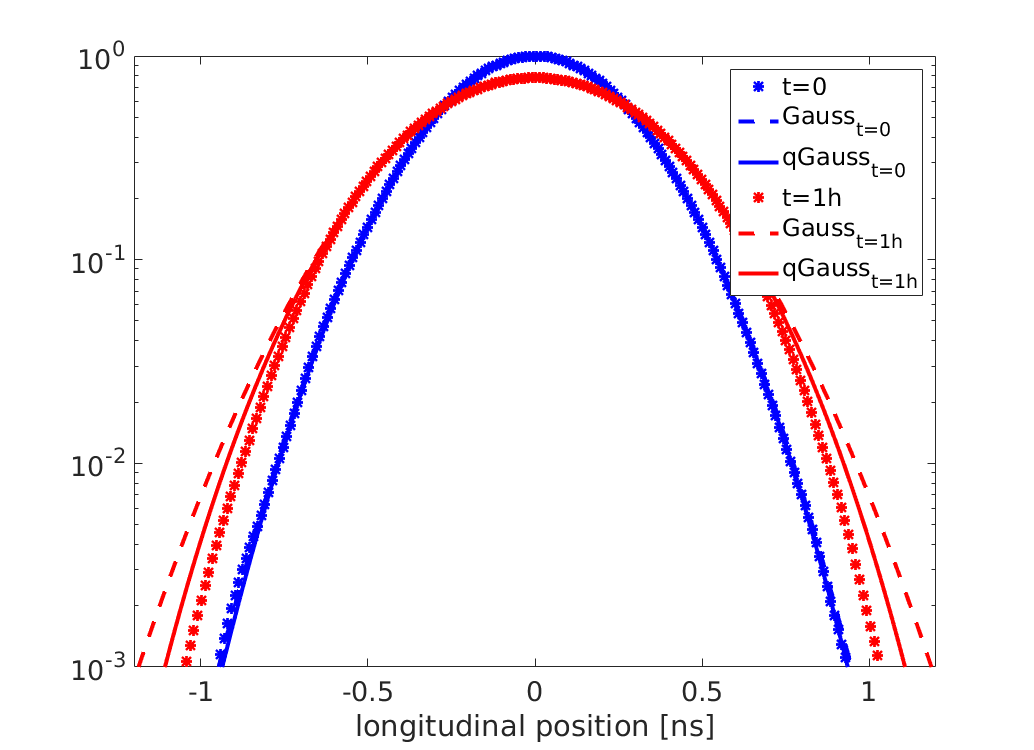}
			\captionsetup{justification=raggedright,singlelinecheck=false}
			\caption{The initial and final (after 1~h) distributions in the horizontal (left), vertical (middle) and longitudinal (right) plane, for the nominal BCMS bunch parameters at injection energy ($450$ GeV), are denoted by blue and red stars, respectively. They are fitted with the Gaussian (dashed line) and the q-Gaussian (solid line) functions.}
			\label{fig:siredistr_nom}
		\end{figure}
		\begin{figure}[h!]
			\centering
			\includegraphics[width=0.325\textwidth]{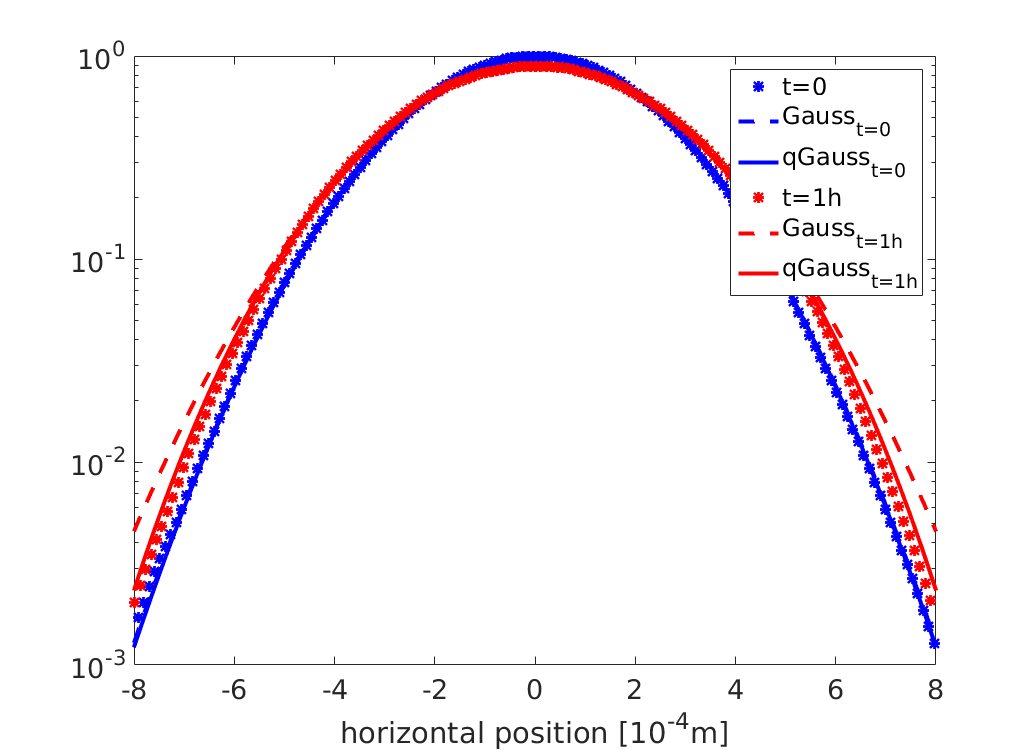}
			\includegraphics[width=0.325\textwidth]{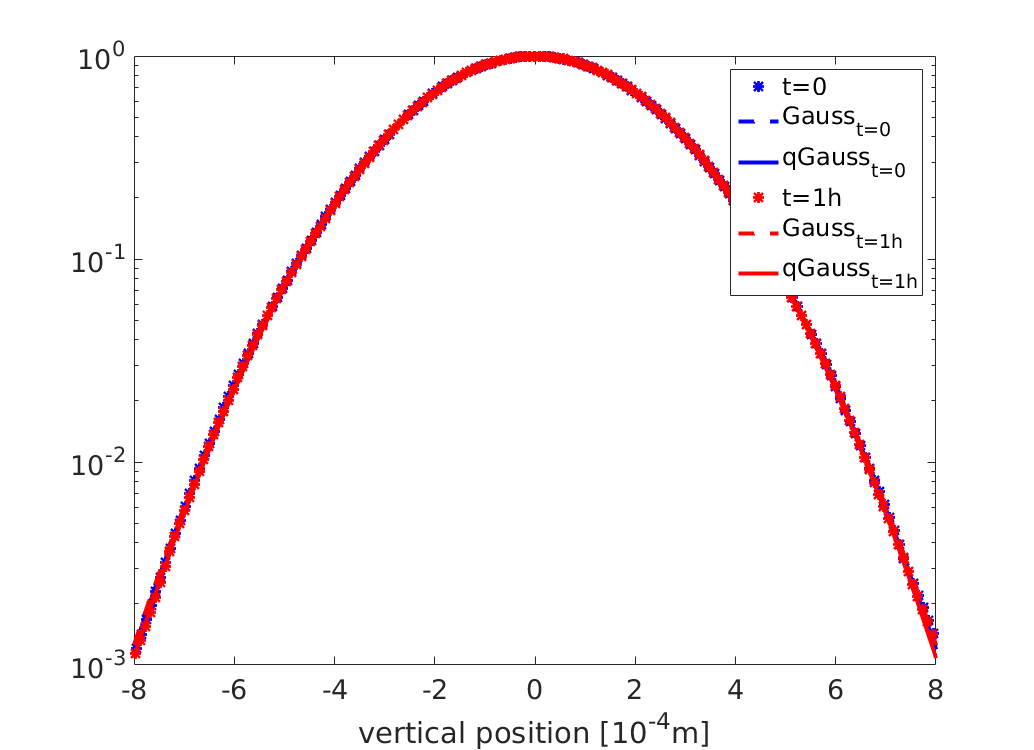}
			\includegraphics[width=0.325\textwidth]{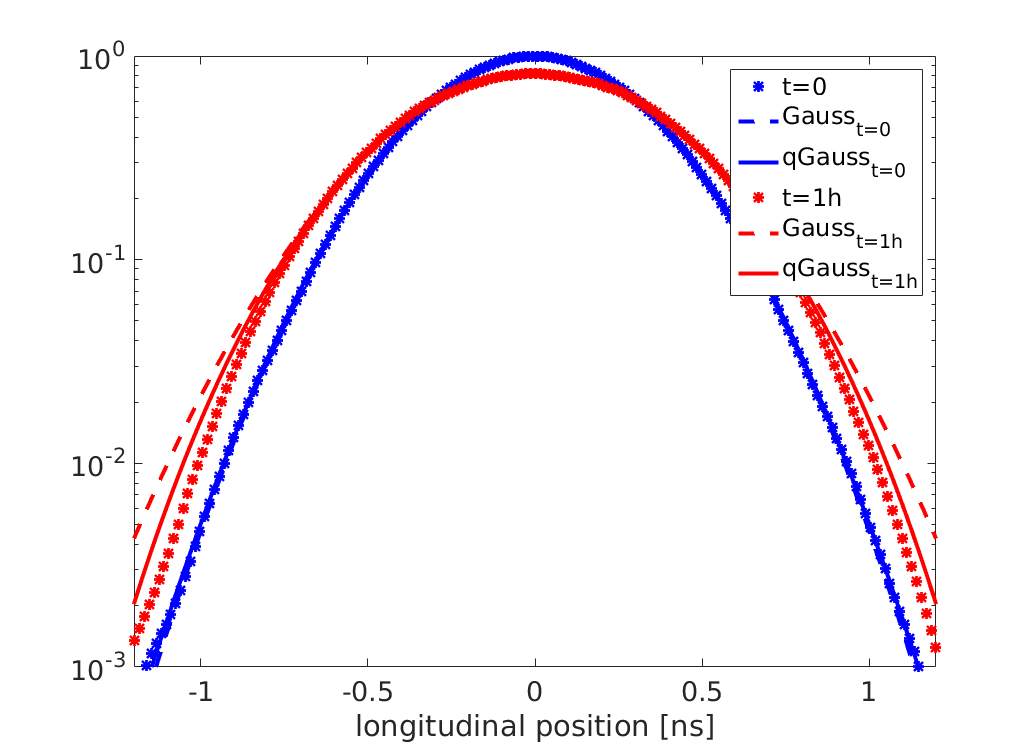}
			\captionsetup{justification=raggedright,singlelinecheck=false}
			\caption{The initial and final (after 1~h) distributions in the horizontal (left), vertical (middle) and longitudinal (right) plane, for the HL-LHC bunch parameters at injection energy ($450$ GeV), are denoted by blue and red stars, respectively. They are fitted with the Gaussian (dashed line) and the q-Gaussian (solid line) functions.}
			\label{fig:siredistr_hilumi}
		\end{figure}	
		\par  In the horizontal and longitudinal plane the IBS effect is dominant, while in the vertical plane, it is minor. 	Even though the SIRE simulation algorithm and the B-M analytical formalism make use of different approaches to calculate the IBS effect (SIRE uses the classical Rutherford cross section   which is closer to the Piwinski formalism), they seem to agree very well during the 1~h time at injection energy. 
		In the longitudinal plane, there is a small difference observed for longer time-spans. Such differences can be explained by the fact that SIRE reshapes the beam distributions, in a self-consistent way, after each collisional process, while the B-M IBS formalism assumes Gaussian beam distributions throughout the calculation. 
		\par The variation of the initially Gaussian particle distributions within 1~h at injection energy is shown in logarithmic scale in Fig.~\ref{fig:siredistr_nom} and Fig.~\ref{fig:siredistr_hilumi} for the nominal BCMS and the HL-LHC case, respectively. The initial and final (after 1~h) distributions in the horizontal (left), vertical (middle) and longitudinal (right) plane, are denoted by blue and red stars, respectively. They are fitted with the Gaussian (dashed line) and the $q$-Gaussian (solid line) functions. The fitting results of the initial and final distributions are presented in Table~\ref{tab:fitparams_nomFB} for the nominal BCMS case and in Table~\ref{tab:fitparams_hilumFB} for the  HL-LHC case. 	
		\par As was expected from the results shown in Fig.~\ref{fig:sirebench_nom_fb}-\ref{fig:sirebench_hilumi_fb} concerning the IBS growth, the horizontal and longitudinal rms beam sizes get larger as time evolves, while the vertical one does not change. The vertical distributions remain Gaussian since  $q\approx1$. For both the nominal and the HL-LHC case, the $q$ parameter of the horizontal and longitudinal distributions is decreased. This can be explained by the fact that, due to IBS,  the core of the distributions is blown up in such a way that it covers up the initially Gaussian tails of the input distributions, which remain less affected. In the longitudinal plane the decrease in $q$ is more significant for the HL-LHC case. This indicates that the stronger IBS is, the more the core is blown up. 
		Since for a light tailed distribution ($q<1$) the Gaussian fit overestimates the rms value, the resulted beam sizes are slightly larger than in the case of the $q$-Gaussian fit. Comparing the root mean square error (RMSE) values of the two fitting functions for the final non-Gaussian bunch profiles shows that the $q$-Gaussian fit is better, in particular for the horizontal plane.
		\begin{table}[h!]
			\caption{Initial and final (after 1~h) fit results for the horizontal, vertical and longitudinal bunch profiles shown in Fig.~\ref{fig:siredistr_nom}, for the nominal BCMS parameters case at injection energy ($450$ GeV).}
			\label{tab:fitparams_nomFB}
			\centering
			\begin{tabular}{lcccccccc}\hline\hline
				\multirow{2}{*}{\textbf{Fit Parameters}} & \multicolumn{2}{c}{ }		& \multicolumn{2}{c}{\textbf{Horizontal distribution}}	& \multicolumn{2}{c}{\textbf{Vertical distribution}} & \multicolumn{2}{c}{\textbf{Longitudinal distribution}} 		\\
				& &	&  $Initial$     	& $Final$ 	&  $Initial$     	& $Final$ &  $Initial$     	& $Final$ \\\hline
				\multirow{2}{*}{Gaussian}	&$\sigma_{rms}\pm 10^{-3} $ &	
				& $0.19~[mm]$		& $0.22~[mm]$	& $0.19~[mm]$		& $0.19~[mm]$ & $0.25~[ns]$	 & $0.33~[ns]$	\\
				&$RMSE~[10^{-3}]$ &							
				& 1		& 14		& 1 	& 1 & 1	 & 10
				\\\hline
				\multirow{3}{*}{q-Gaussian}	&$\sigma_{rms}\pm10^{-3}$ &	
				& $0.19~[mm]$		& $0.21~[mm]$	& $0.19~[mm]$		& $0.19~[mm]$ & $0.25~[ns]$	 & $0.32~[ns]$	\\
				&$q\pm dq$ &						
				& $1.024\pm0.003$	 & $0.893\pm0.002$ & $0.970\pm0.007$	& $0.967\pm0.006$ & $0.992\pm0.002$	 & $0.941\pm0.001$\\
				&$RMSE~[10^{-3}]$ &							
				& 1	& 1	& 1	 & 1 & 1	 & 6
				\\\hline\hline
			\end{tabular}
		\end{table} 
		\begin{table}[h!]
			\caption{Initial and final (after 1~h) fitting results for the horizontal, vertical and longitudinal bunch profiles shown in Fig.~\ref{fig:siredistr_hilumi}, for the HL-LHC parameters case at injection energy ($450$ GeV).}
			\label{tab:fitparams_hilumFB}
			\centering
			\begin{tabular}{lcccccccc}\hline\hline
				\multirow{2}{*}{\textbf{Fit Parameters}} & \multicolumn{2}{c}{ }		& \multicolumn{2}{c}{\textbf{Horizontal distribution}}	& \multicolumn{2}{c}{\textbf{Vertical distribution}} & \multicolumn{2}{c}{\textbf{Longitudinal distribution}} 		\\
				& &	&  $Initial$     	& $Final$ 	&  $Initial$     	& $Final$ &  $Initial$     	& $Final$ \\\hline
				\multirow{2}{*}{Gaussian}	&$\sigma_{rms}\pm 10^{-3} $ &	
				& $0.22~[mm]$		& $0.25~[mm]$	& $0.22~[mm]$		& $0.22~[mm]$ & $0.30~[ns]$	 & $0.37~[ns]$	\\
				&$RMSE~[10^{-3}]$ &							
				& 1		& 14		& 1 	& 1 & 3	 & 13
				\\\hline
				\multirow{3}{*}{q-Gaussian}	&$\sigma_{rms}\pm 10^{-3}$ &	
				& $0.22~[mm]$		& $0.24~[mm]$	& $0.22~[mm]$		& $0.22~[mm]$ & $0.30~[ns]$	 & $0.36~[ns]$	\\
				&$q\pm dq$ &						
				& $0.992\pm0.003$	& $0.891\pm0.004$ & $0.995\pm0.003$	 & $0.987\pm0.003$ & $1.019\pm0.005$	 & $0.885\pm0.001$\\
				&$RMSE~[10^{-3}]$ &							
				& 1	& 1	& 1	 & 1 & 3 & 4
				\\\hline\hline
			\end{tabular}
		\end{table} 

		\subsubsection{At the LHC collision energy (6.5~TeV)}
		\par Since at collision energy IBS becomes weaker and synchrotron radiation starts playing an important role, it is the interplay between these effects that determines the evolution of the bunch characteristics. In this respect, for the benchmarking of the B-M IBS theoretical model with SIRE at collision energy, apart from the IBS, the radiation effects (synchrotron radiation and quantum excitation) are also taken into account. It should be mentioned that for the results presented in the following plots the intensity is assumed to be constant (which is actually true for the few non-colliding bunches, during physics fills). 
		\par
\begin{figure}[h]
			\centering
			\includegraphics[width=0.325\textwidth]{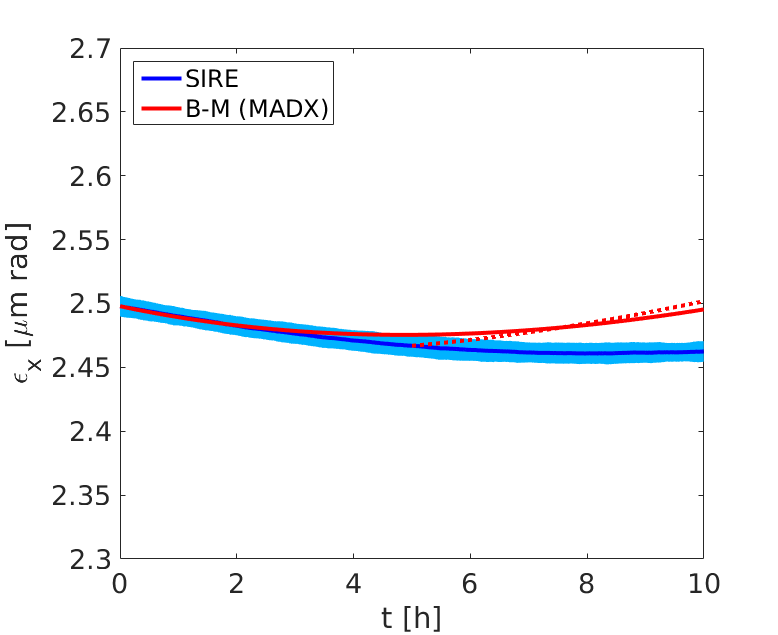}
			\includegraphics[width=0.325\textwidth]{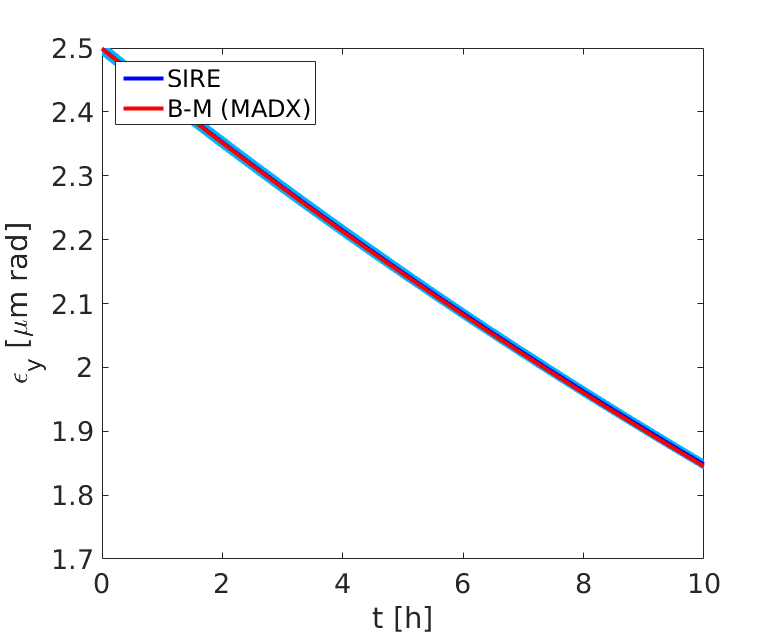}
			\includegraphics[width=0.325\textwidth]{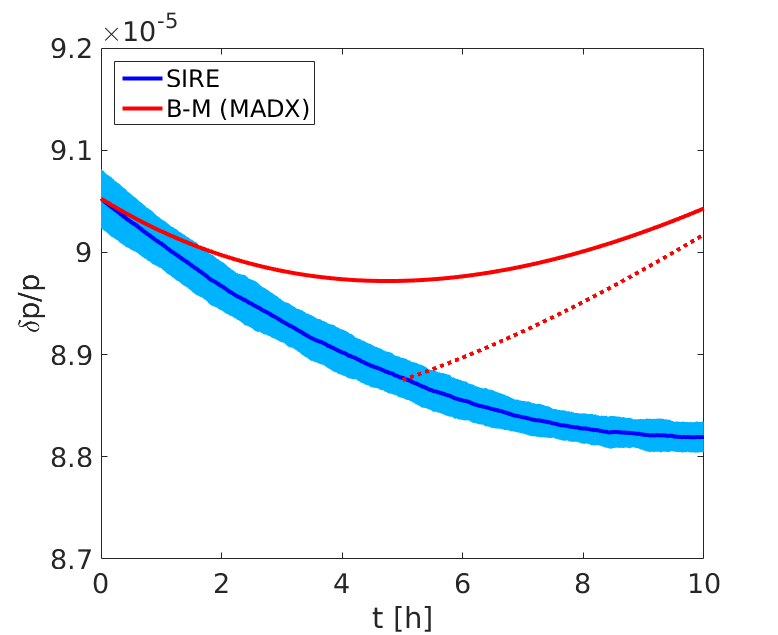}
			\captionsetup{justification=raggedright,singlelinecheck=false}
			\caption{The evolution of the horizontal (left) and vertical (middle) emittance and energy spread (right) due to IBS and radiation effects, in a time period of 10~h at the collision energy of the LHC  ($6.5$ TeV) for the nominal BCMS parameters, as computed by the SIRE code (blue line) and the Bjorken-Mtingwa analytical formalism in MAD-X (red line).}
			\label{fig:sirebench_nom_ft}
		\end{figure}

\begin{figure}[h]
			\centering
			\includegraphics[width=0.325\textwidth]{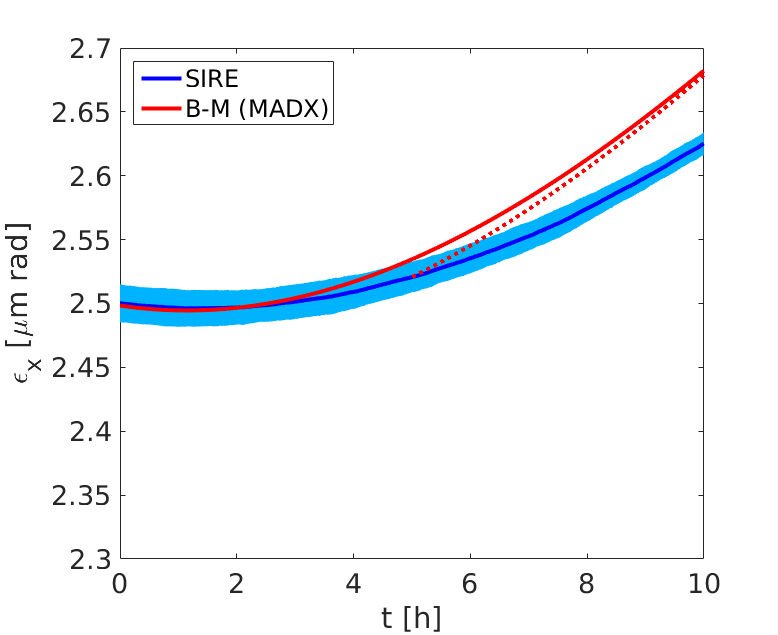}
			\includegraphics[width=0.325\textwidth]{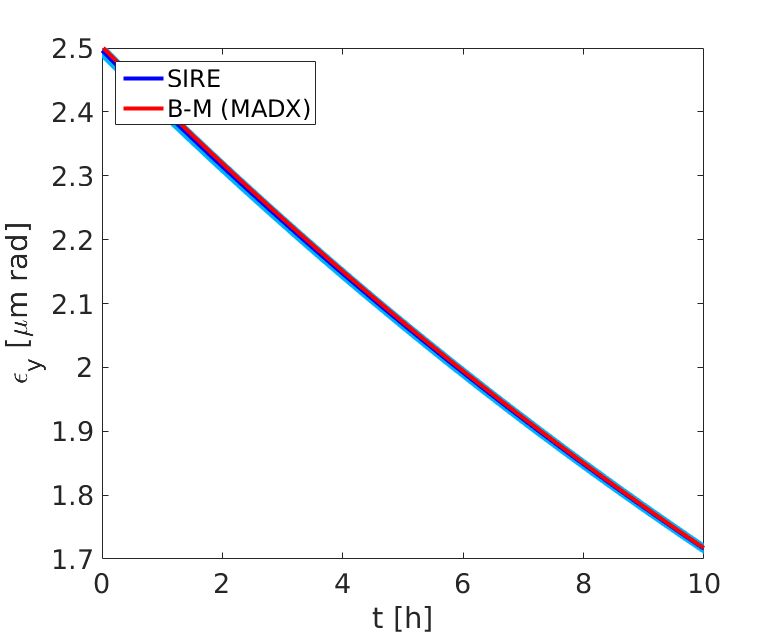}
			\includegraphics[width=0.325\textwidth]{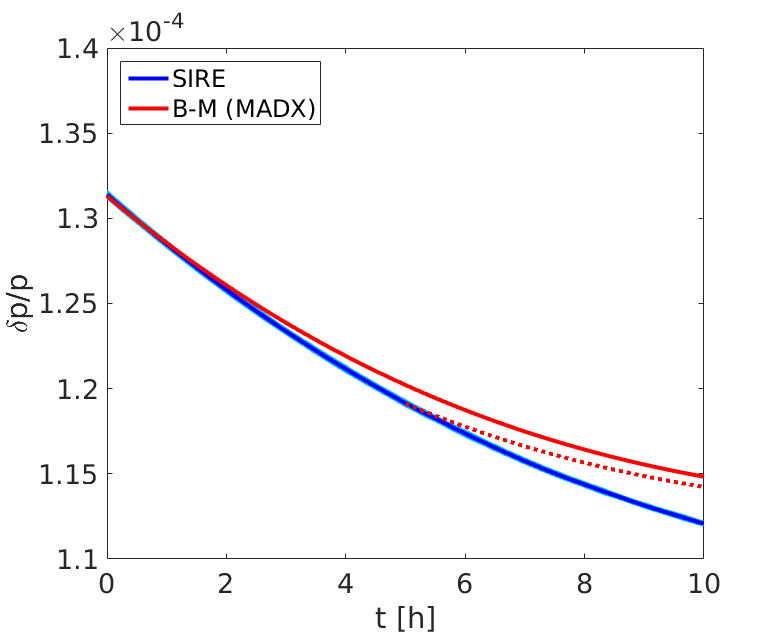}
			\captionsetup{justification=raggedright,singlelinecheck=false}
			\caption{The evolution of the horizontal (left) and vertical (middle) emittance and energy spread (right) due to IBS and radiation effects, in a time period of 10~h at the collision energy of the LHC  ($7$ TeV) for the HL-LHC parameters, as computed by the SIRE (blue line) and the Bjorken-Mtingwa analytical formalism in MAD-X (red line).}
			\label{fig:sirebench_hilumi_ft}
		\end{figure}

 Figure~\ref{fig:sirebench_nom_ft} shows the horizontal emittance (left), the vertical emittance (middle) and energy spread (right) evolution after 10~h at collision energy for the nominal BCMS case, while Fig.~\ref{fig:sirebench_hilumi_ft} shows the evolutions for the HL-LHC parameters. The red and the blue lines correspond to the analytical calculations of the MAD-X~\cite{ref:MADX} IBS routine (based on the B-M formalism) and to the SIRE results, respectively.  The two standard deviation error-bars for the simulation results are plotted in light blue. Table~\ref{tab:IBSgrowths_ft} summarizes the variation of the transverse emittances and energy spread during 10~h at the collision energy of the LHC, for the nominal BCMS and HL-LHC parameters, as computed by the SIRE code and the B-M analytical formalism in MAD-X. 
		\begin{table}[h!]
			\caption{Variation of the transverse emittances and energy spread during 10~h at FT energy.}
			\label{tab:IBSgrowths_ft}
			\centering
			\begin{tabular}{lcccc}\hline\hline
				\multirow{2}{*}{\textbf{IBS  growths}}		& \multicolumn{2}{c}{\textbf{Nominal (BCMS)}}	& \multicolumn{2}{c}{\textbf{HL-LHC}} 
				\\
				&  $MAD-X$     	& $SIRE$ 	&  $MAD-X$     	& $SIRE$ \\\hline
				$d\epsilon_x/\epsilon_{x0}$ [\%]		& -0.1 	& -1.4	& 7.4		& 5.0	\\
				$d\epsilon_y/\epsilon_{y0}$ [\%]			& -26.2		& -26.1		& -31.4 	& -31.2		\\
				$d\sigma_l/\sigma_{l0}$	   [\%]		& -0.1	& -2.6	& -12.6	 & -14.7	\\\hline\hline
			\end{tabular}
		\end{table} 
		\par After a few hours at collisions, the B-M analytical formalism and the simulations start differentiating. In order to understand whether these differences are explained by the fact that SIRE reshapes the beam distributions after each collisional process and the B-M IBS formalism assumes always Gaussian beam distributions, the bunch parameters given by SIRE at 5~h are used as input for the IBS and synchrotron radiation calculations in MAD-X (Gaussian bunches). The red dotted lines in Fig.~\ref{fig:sirebench_nom_ft} and Fig.~\ref{fig:sirebench_hilumi_ft} represent the results of these tests. Even if giving as input to MAD-X exactly the same bunch parameters as in SIRE, there is clear divergence of the MAD-X results (red dotted lines) with SIRE right after the 5~h at collisions. This divergence is much larger than the one observed during the first hours at collisions. After 5~h at collision energy, the beam in SIRE has been reshaped enough so that IBS and radiation processes act differently as compared to Gaussian MAD-X distributions. Consequently, the differences observed between the B-M analytical formalism and the simulations are expected because MAD-X assumes always Gaussian distribution, in contrast to SIRE that takes into account the variation of the bunch shape throughout the calculation.
		\par Due to the fact that the IBS effect is minor in the vertical plane, the strong synchrotron radiation damping mechanism leads to a clear reduction of the vertical emittance. However, the variation of the horizontal emittance and energy spread is determined by the interplay of IBS growth with synchrotron radiation damping. For the nominal BCMS parameters, these variations are very small after 10~h at collision energy (Table~\ref{tab:IBSgrowths_ft}). For the HL-LHC case, having the same initial horizontal emittance but the double bunch population compared to the nominal BCMS parameters (Table~\ref{tab:params}), the IBS effect prevails over synchrotron radiation in the horizontal plane after almost 3~h (Fig.~\ref{fig:sirebench_hilumi_ft} (left)). As can be seen in Fig.~\ref{fig:sirebench_hilumi_ft} (right) this in not the case for the longitudinal plane, for which the initial bunch length of 1.2~ns compared to the 1~ns in the nominal case, renders IBS weaker than synchrotron radiation and, results in the decrease of the energy spread.
		\begin{figure}[h]
			\centering
			\includegraphics[width=0.325\textwidth]{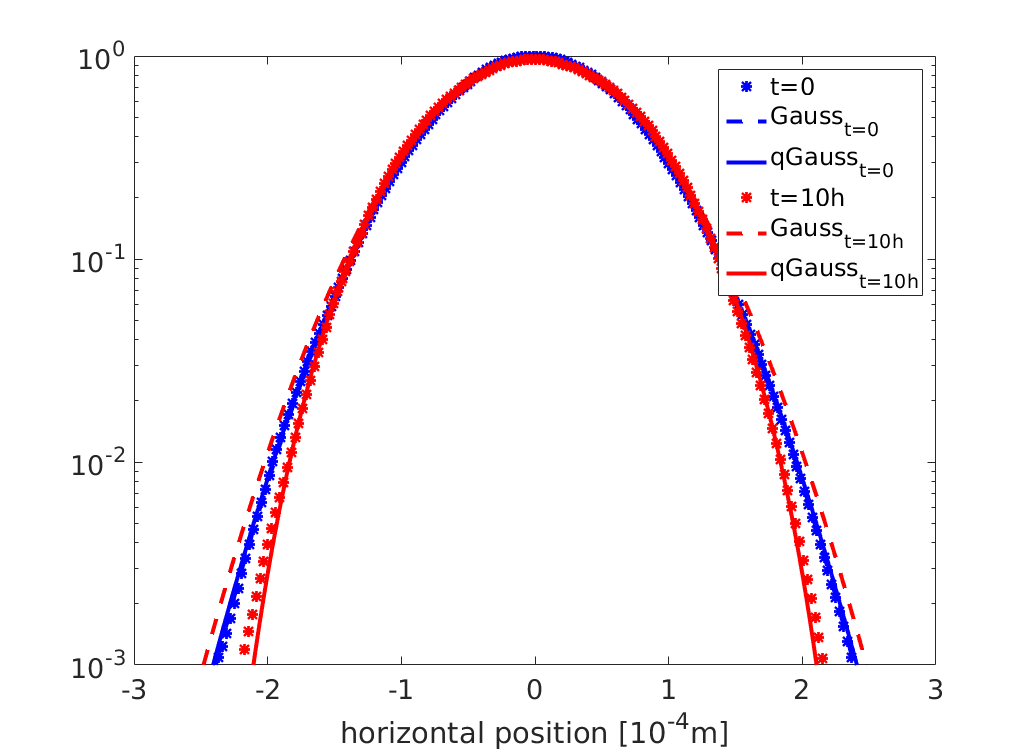}
			\includegraphics[width=0.325\textwidth]{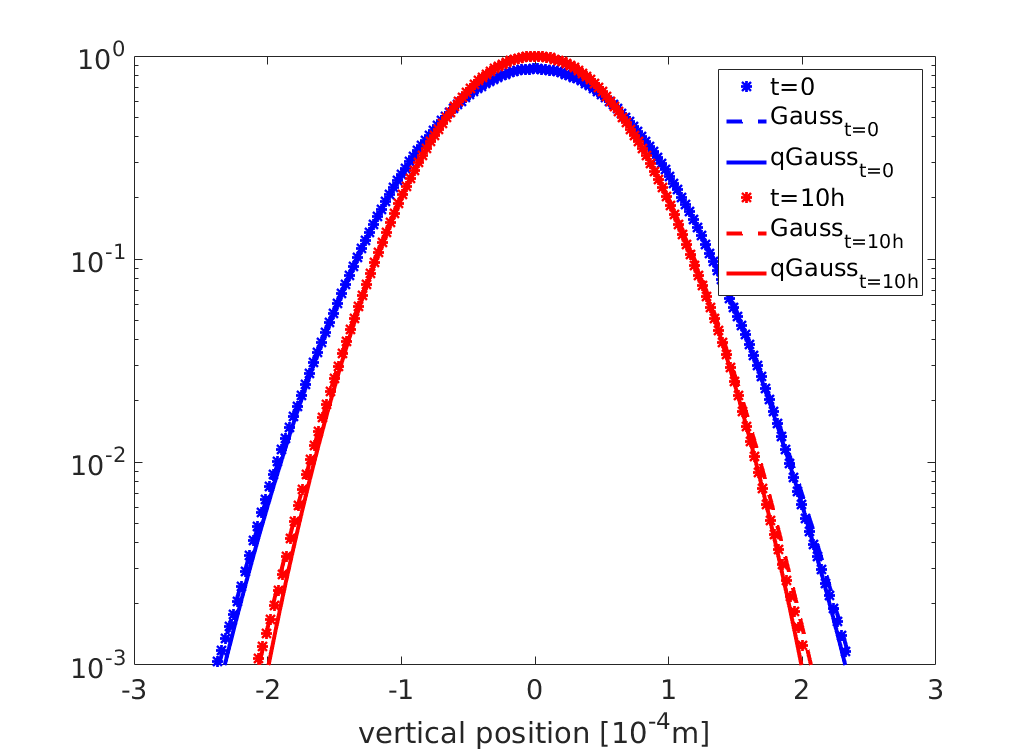}
			\includegraphics[width=0.325\textwidth]{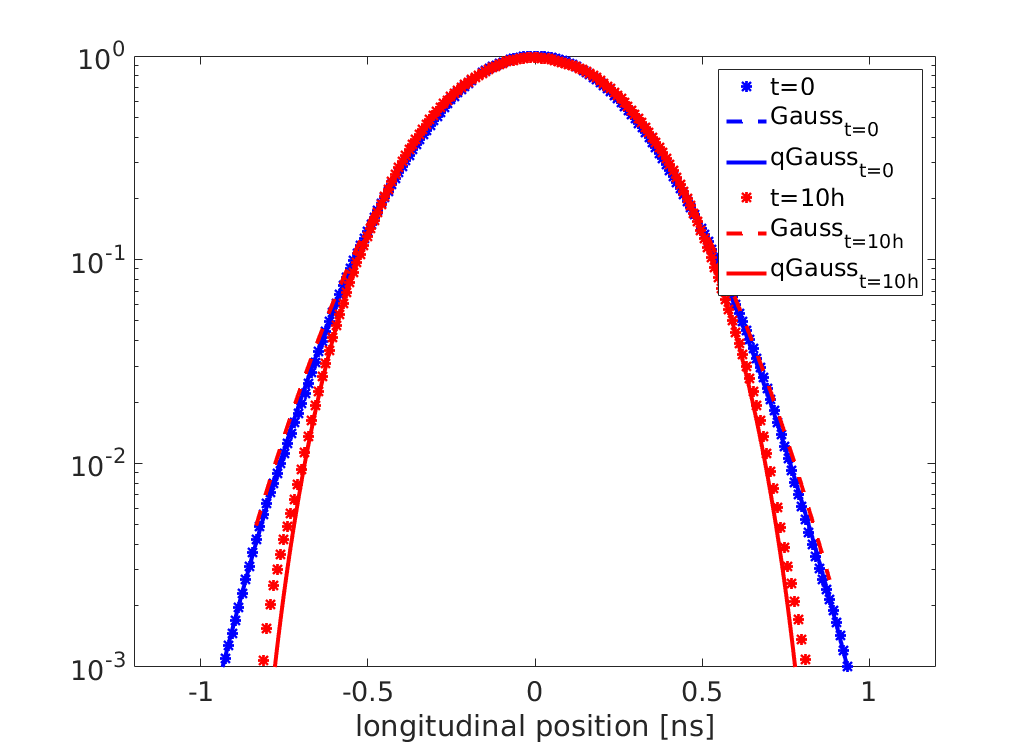}
			\captionsetup{justification=raggedright,singlelinecheck=false}
			\caption{The initial and final (after 10~h) distributions in the horizontal (left), vertical (middle) and longitudinal (right) plane, for the nominal BCMS bunch parameters at collision energy ($6.5$ TeV), are denoted by blue and red stars, respectively. They are fitted with the Gaussian (dashed line) and the q-Gaussian (solid line) functions.}
			\label{fig:siredistr_nom_ft}
		\end{figure}
		\begin{figure}[h!]
			\centering
			\includegraphics[width=0.325\textwidth]{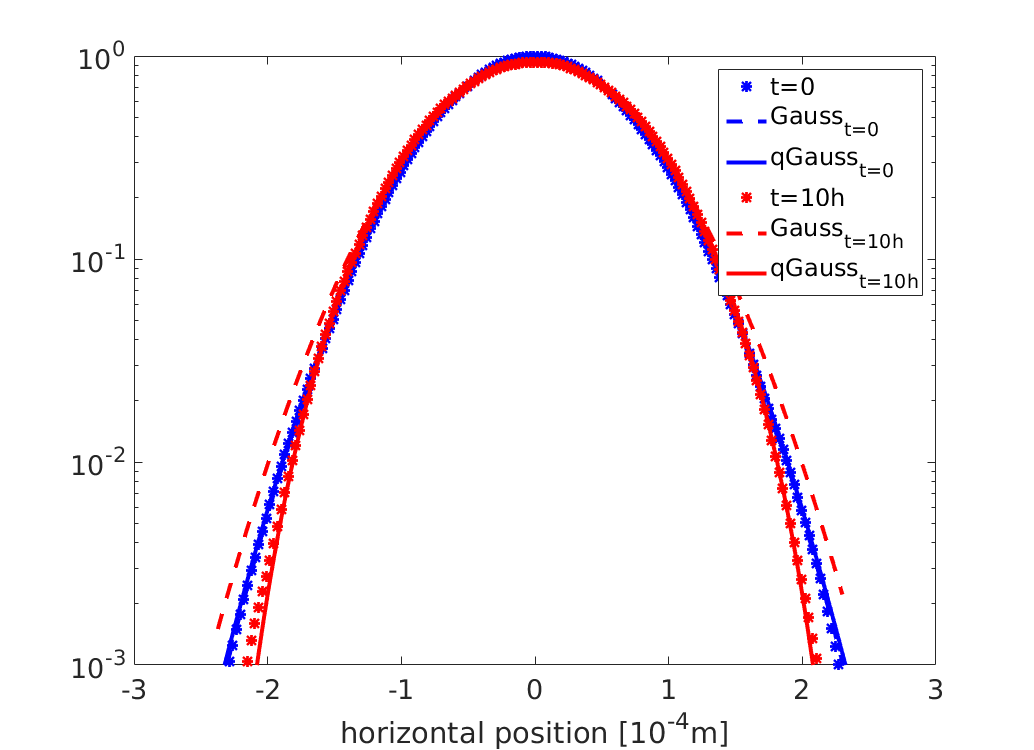}
			\includegraphics[width=0.325\textwidth]{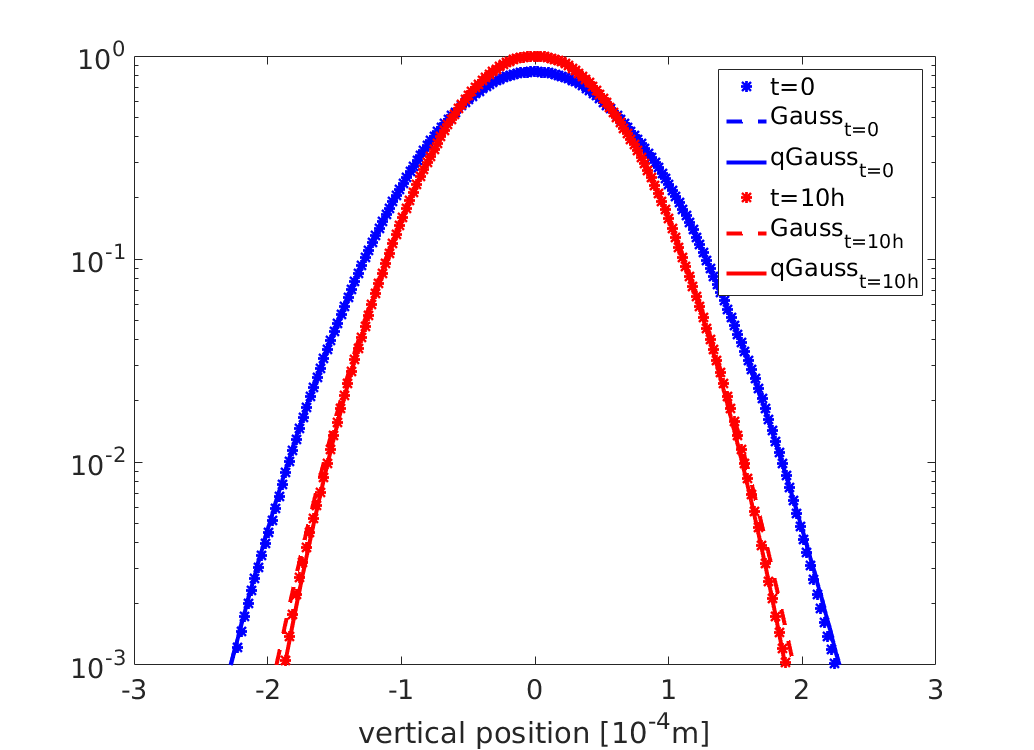}
			\includegraphics[width=0.325\textwidth]{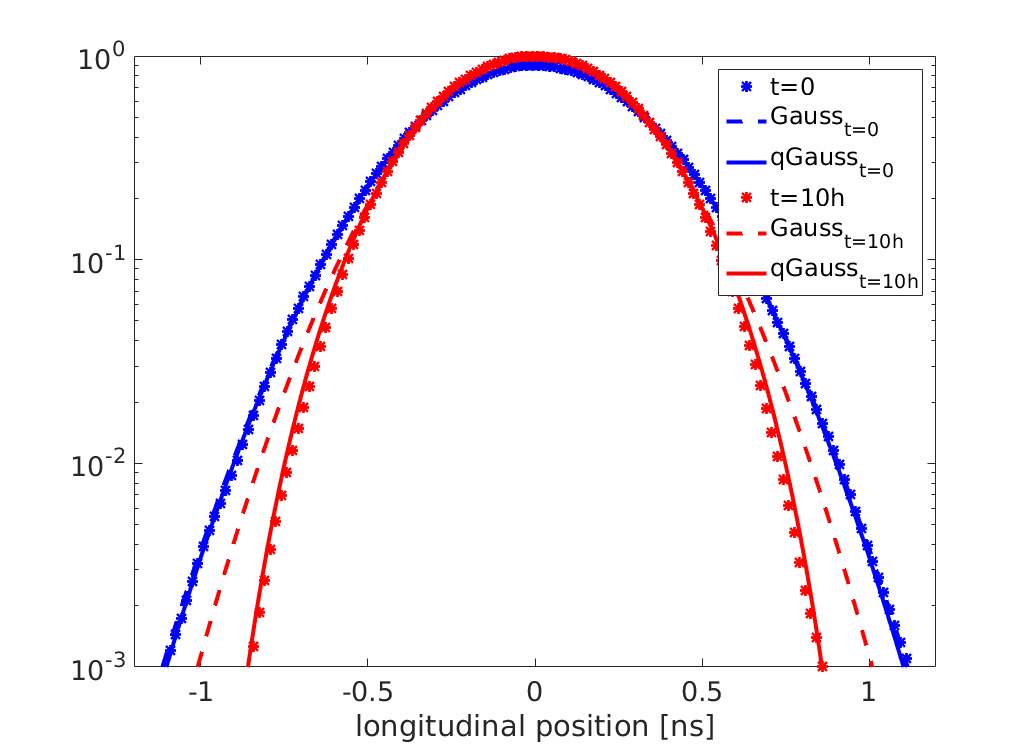}
			\captionsetup{justification=raggedright,singlelinecheck=false}
			\caption{The initial and final (after 10~h) distributions in the horizontal (left), vertical (middle) and longitudinal (right) plane, for the HL-LHC bunch parameters at collision energy ($7$ TeV), are denoted by blue and red stars, respectively. They are fitted with the Gaussian (dashed line) and the q-Gaussian (solid line) functions.}
			\label{fig:siredistr_hilumi_ft}
		\end{figure} 
		\begin{table}[h!]
			\caption{Initial and final (after 10~h) fit results for the horizontal, vertical and longitudinal bunch profiles shown in Fig.~\ref{fig:siredistr_nom_ft}, for the nominal BCMS parameters case at collision energy ($6.5$ TeV).}
			\label{tab:fitparams_nomFT}
			\centering
			\begin{tabular}{lcccccccc}\hline\hline
				\multirow{2}{*}{\textbf{Fit Parameters}} & \multicolumn{2}{c}{ }		& \multicolumn{2}{c}{\textbf{Horizontal distribution}}	& \multicolumn{2}{c}{\textbf{Vertical distribution}} & \multicolumn{2}{c}{\textbf{Longitudinal distribution}} 		\\
				& &	&  $Initial$     	& $Final$ 	&  $Initial$     	& $Final$ &  $Initial$     	& $Final$ \\\hline
				\multirow{2}{*}{Gaussian}	&$\sigma_{rms}\pm 10^{-4} $ &	
				& $0.064~[mm]$		& $0.067~[mm]$	& $0.064~[mm]$		& $0.056~[mm]$ & $0.25~[ns]$	 & $0.26~[ns]$	\\
				&$RMSE~[10^{-3}]$ &							
				& 1		& 25		& 1 	& 1 & 1	 & 30
				\\\hline
				\multirow{3}{*}{q-Gaussian}	&$\sigma_{rms}\pm10^{-4}$ &	
				& $0.064~[mm]$		& $0.064~[mm]$	& $0.064~[mm]$		& $0.055~[mm]$ & $0.25~[ns]$	 & $0.24~[ns]$	\\
				&$q\pm dq$ &						
				& $1.004\pm0.003$	 & $0.856\pm0.005$ & $0.982\pm0.004$	& $0.971\pm0.004$ & $1.007\pm0.004$	 & $0.830\pm0.006$\\
				&$RMSE~[10^{-3}]$ &							
				& 1	& 1	& 1	 & 1 & 1	 & 1
				\\\hline\hline
			\end{tabular}
		\end{table} 
		
		\begin{table}[h!]
			\caption{Initial and final (after 10~h) fitting results for the horizontal, vertical and longitudinal bunch profiles shown in Fig.~\ref{fig:siredistr_hilumi_ft}, for the HL-LHC parameters case at collision energy ($7$ TeV).}
			\label{tab:fitparams_hilumFT}
			\centering
			\begin{tabular}{lcccccccc}\hline\hline
				\multirow{2}{*}{\textbf{Fit Parameters}} & \multicolumn{2}{c}{ }		& \multicolumn{2}{c}{\textbf{Horizontal distribution}}	& \multicolumn{2}{c}{\textbf{Vertical distribution}} & \multicolumn{2}{c}{\textbf{Longitudinal distribution}} 		\\
				& &	&  $Initial$     	& $Final$ 	&  $Initial$     	& $Final$ &  $Initial$     	& $Final$ \\\hline
				\multirow{2}{*}{Gaussian}	&$\sigma_{rms}\pm 10^{-4} $ &	
				& $0.062~[mm]$		& $0.067~[mm]$	& $0.062~[mm]$		& $0.052~[mm]$ & $0.30~[ns]$	 & $0.28~[ns]$	\\
				&$RMSE~[10^{-3}]$ &							
				& 1		& 27		& 2 	& 2 & 2	 & 17
				\\\hline
				\multirow{3}{*}{q-Gaussian}	&$\sigma_{rms}\pm10^{-4}$ &	
				& $0.062~[mm]$		& $0.063~[mm]$	& $0.062~[mm]$		& $0.052~[mm]$ & $0.30~[ns]$	 & $0.27~[ns]$	\\
				&$q\pm dq$ &						
				& $1.005\pm0.004$	 & $0.852\pm0.004$ & $0.991\pm0.005$	& $0.977\pm0.005$ & $0.990\pm0.003$	 & $0.825\pm0.001$\\
				&$RMSE~[10^{-3}]$ &							
				& 1	& 1	& 1	 & 1 & 1	 & 1
				\\\hline\hline
			\end{tabular}
		\end{table}   
		\par The evolution of the initially Gaussian (in all planes) particle distributions within 10~h at collision energy is shown in logarithmic scale in Fig.~\ref{fig:siredistr_nom_ft} and Fig.~\ref{fig:siredistr_hilumi_ft} for the nominal BCMS and the HL-LHC case, respectively. The initial and final (after 10~h) distributions in the horizontal (left), vertical (middle) and longitudinal (right) plane, are denoted by blue and red stars, respectively. They are fitted with the Gaussian (dashed line) and the $q$-Gaussian (solid line) functions. The fitting results of the initial and final distributions are presented in Table~\ref{tab:fitparams_nomFT} for the nominal BCMS case and in Table~\ref{tab:fitparams_hilumFT} for the  HL-LHC case.  The RMSE values of the two fitting functions show that when the final bunch profiles are strongly non-Gaussian, the q-Gaussian fitting results should be considered. In this respect, the evolution of the particle distributions in all planes for the nominal BCMS and HL-LHC cases is discussed based on the $q$-Gaussian results.
		\par The horizontal beam sizes do not change after 10~h at collision energy because the blow up caused by IBS is balanced out by the synchrotron radiation damping. However, there is a transformation of the horizontal distributions' shape for which the tails become less populated ($q<1$). In the longitudinal plane both the beam size and the $q$ parameter are reduced, meaning that synchrotron radiation prevails over IBS and the core is blown up due to IBS giving underpopulated tails. In the vertical plane, the dominant synchrotron radiation damping results in a smaller beam size without changing much the formation of the tails, so the distribution remains Gaussian. 

	\section{Bunch profile measurements in the LHC}
	\label{measured}
		\par The transverse diagnostic instruments for measuring the bunch profiles in the LHC are the betatron matching monitor~\cite{ref:burger}, the Beam Gas Ionization (BGI) monitor~\cite{ref:sapinski}, the Beam Gas Vertex (BGV) monitor, the Beam Wire Scanners (WS)~\cite{ref:bosser} and the Beam Synchrotron Light Monitor (BSRT)~\cite{ref:meot}.  Compatibly with high intensity and high energy operation, the BSRT is the only instrument offering non-invasive, continuous and bunch-by-bunch measurements of the LHC beams. The BSRT is calibrated with respect to the WS during dedicated low beam intensity runs.~\footnote{The WS can measure the emittance throughout the full LHC machine cycle including the energy ramp, provided that the total intensity in the machine is limited to 240 nominal bunches at 450 GeV and 12 nominal bunches at 6.5~TeV.} The LHC is equipped with two synchrotron radiation monitors (one per beam) used to characterize the transverse and longitudinal beam distributions. Due to the significant diffraction patterns at the tails of the BSRT transverse distributions there is no clear picture about the shape of the tails and so, they are often assumed to be Gaussian.
		\par A parameter that is generally used to measure the longitudinal emittance in circular accelerators is the bunch length. The bunch length is given by  the projection of the distribution function on the phase axis, which is known as the bunch profile or line density.  It is operationally measured by the LHC Beam Quality Monitor (BQM)~\cite{ref:papotti} which uses a wall current monitor pick-up (WCM)~\cite{ref:wcm} to acquire the longitudinal profiles. Additionally, the longitudinal synchrotron radiation monitor (BSRL)~\cite{ref:bsrl}, which uses the same synchrotron light source as the BSRT, continuously measures the longitudinal distribution of charges in the beams.
		The scopes connected to the WCM pick-ups can acquire longitudinal bunch profiles of both beams during a full LHC cycle. 

	\subsection{Comparison between experimental data, the SIRE and the B-M analytical formalism, at the LHC collision energy (6.5~TeV)}
	
	The longitudinal bunch manipulations performed during the ramp to avoid instabilities due to the loss of Landau damping~\cite{ref:Baudrenghien}, produce bunches that arrive at collision energy with a clearly non-Gaussian longitudinal shape. In addition, the transfer functions of the pickups and cables were measured and are used for deconvolution~\cite{ref:juan}, resulting in some cases in tails which are asymmetric or have ripples. 
	\begin{figure}[h]
		\centering
		\includegraphics[width=0.4\textwidth]{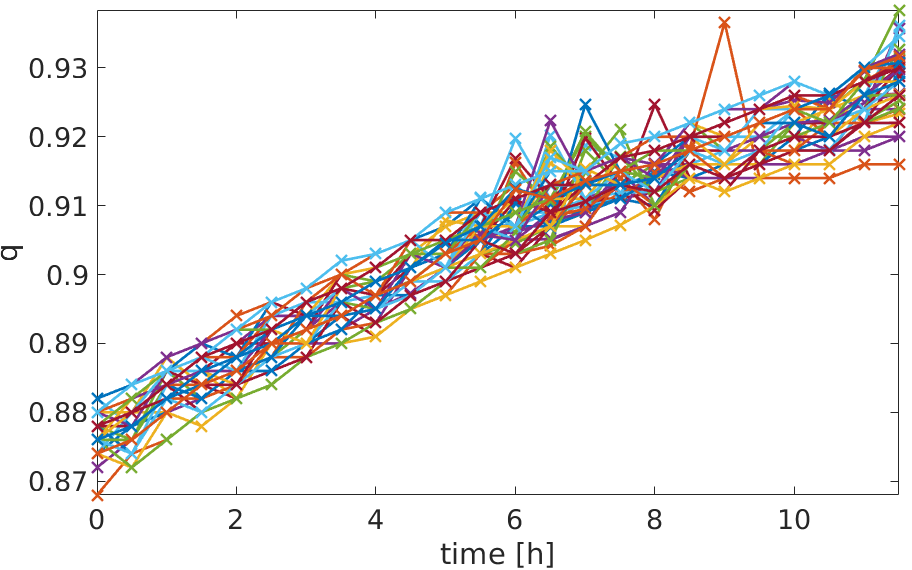}
		\caption{The evolution of the $q$ parameter during 11.5~h at collisions (6.5~TeV), for a train of bunches in the longitudinal plane.}
		\label{fig:qevol_12h}
	\end{figure}
	By assuming that these profiles are Gaussian may lead in underestimating or overestimating the actual bunch length. For the studies presented in this paper, these profiles are fitted using the $q$-Gaussian function. An example showing the evolution of the $q$ parameter for the longitudinal profile of a bunch train during 11.5~h at collisions (6.5~TeV) in the LHC is presented in Fig.~\ref{fig:qevol_12h}. It is clear that with such $q$ parameter values, corresponding to non-Gaussian tails, the rms beam size cannot be accurately estimated by using the Gaussian function.  
	\begin{figure}[h]
		\centering
		\includegraphics[width=0.45\textwidth]{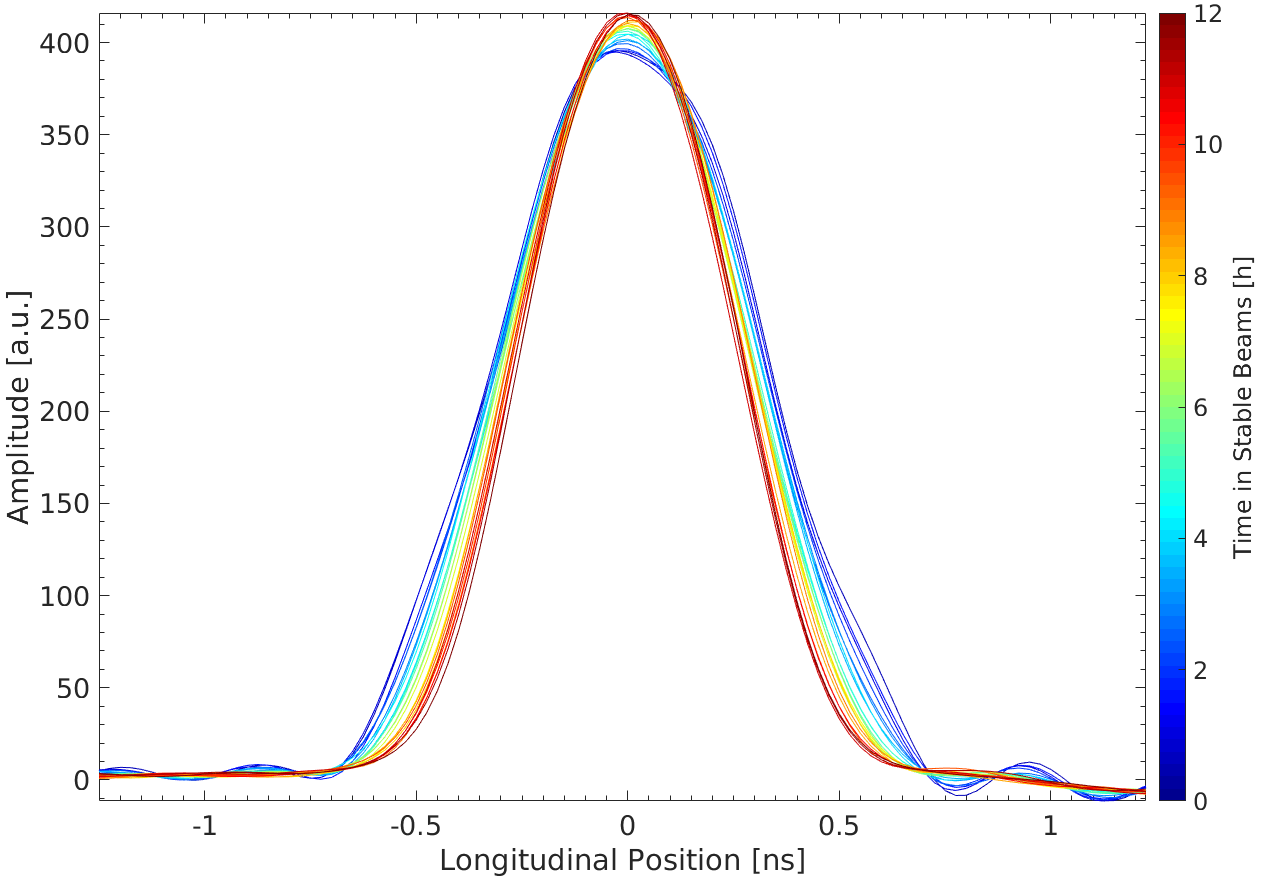}
		\caption{The evolution of a longitudinal bunch profile during 11.5~h at collisions (6.5~TeV).}
		\label{fig:distribevol_12h}
	\end{figure}
	The increase of the $q$ parameter means that the longitudinal distributions with the underpopulated tails ($q<1$) at the start of collisions, become more Gaussian ($q\rightarrow1$) as time evolves. This is a general statement  that can be made for the longitudinal distribution observed at the collision energy of the LHC. The evolution of the longitudinal particle distribution of a single bunch that is picked out of the train of bunches is shown in Fig.~\ref{fig:distribevol_12h} for the time period of 11.5~h. The initial bunch profile (plotted in blue) is fitted with the Gaussian and the q-Gaussian functions that give different rms beam sizes because of the dependence of the standard deviation on the $q$ parameter (Eq.~\eqref{eq:sigma_qGauss}). The fitting results are used to generate a Gaussian and a q-Gaussian distribution to be tracked in SIRE in order to compare the experimental observations with the results of the code. 
	\begin{figure}[h]
		\centering
		\includegraphics[width=0.37\textwidth]{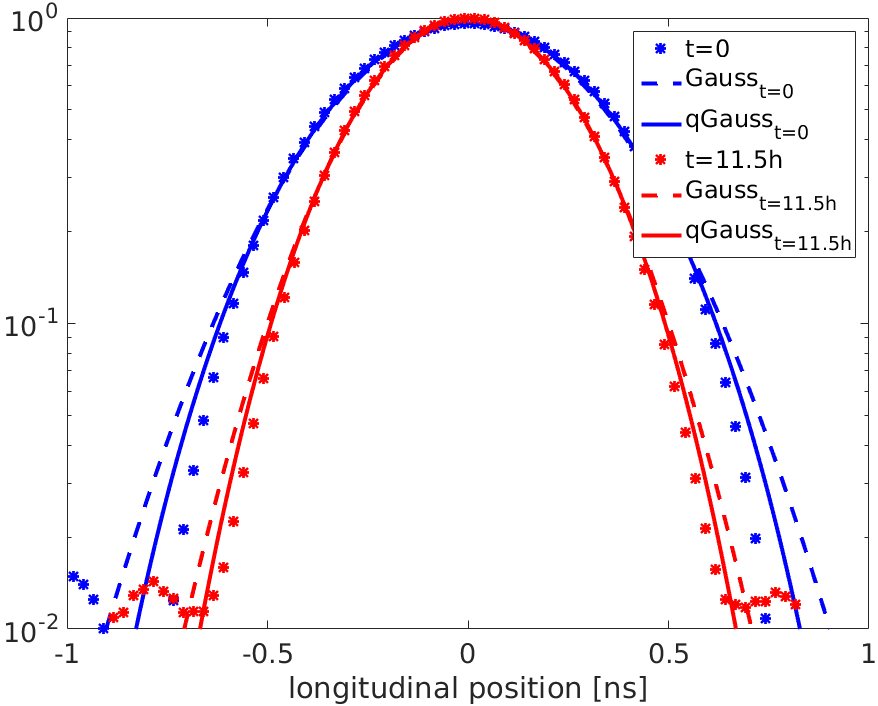}
		~~~~~	
		\includegraphics[width=0.37\textwidth]{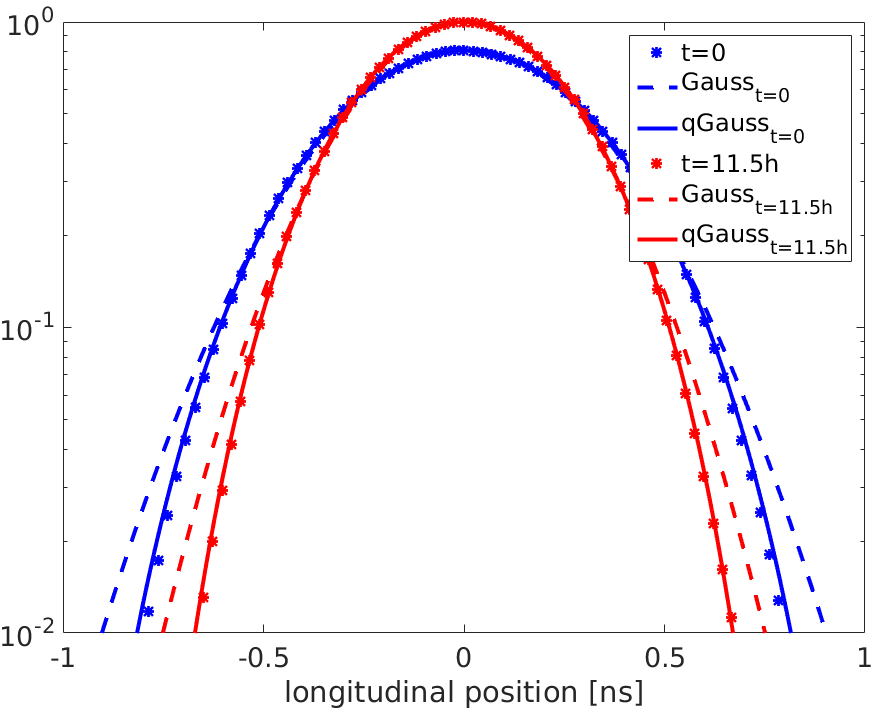}
		\captionsetup{justification=raggedright,singlelinecheck=false}
		\caption{The initial (at the start of collisions) and the final (after
			11.5 h) longitudinal bunch profiles as observed in the LHC (left) and as calculated by the SIRE (right), in logarithmic scale, are denoted by blue and red stars, respectively. They are fitted with the	Gaussian (dashed line) and the $q$-Gaussian (solid line) functions.}
		\label{fig:distrib_12h}
	\end{figure}
	\begin{table}[h]
		\caption{Fitting results for the initial (at the start of collisions) and the final (after
			11.5 h) longitudinal bunch distribution shown in Fig.~\ref{fig:distrib_12h}, as was observed in the LHC and as was calculated by the SIRE code.}
		\label{tab:fitparams_12hFT}
		\centering
		\begin{tabular}{lcccccccc}\hline\hline
			\multirow{2}{*}{\textbf{Fit Parameters}} & \multicolumn{2}{c}{ }		& \multicolumn{2}{c}{\textbf{Initial (t=0)}}	& \multicolumn{2}{c}{\textbf{Final (t=11.5~h)}} 		\\
			& &	&  $DATA$     	& $SIRE$ 	&  $DATA$     	& $SIRE$  \\\hline
			\multirow{2}{*}{Gaussian}	&$\sigma_{rms}\pm d\sigma_{rms}~[ns]$ &	
			& $0.299\pm0.003$		& $0.297\pm0.002$	& $0.233\pm0.002$		& $0.237\pm0.002$ 	\\
			&$RMSE~[10^{-3}]$ &							
			& 22		& 19		& 18 	& 20 
			\\\hline
			\multirow{3}{*}{q-Gaussian}	&$\sigma_{rms}\pm d\sigma_{rms}~[ns]$ &	
			& $0.286\pm0.004$		& $0.290\pm0.001$	& $0.227\pm0.002$		& $0.235\pm0.001$ 	\\
			&$q\pm dq$ &						
			& $0.88\pm0.03$	 & $0.85\pm0.01$ & $0.93\pm0.03$	& $0.86\pm0.01$ \\
			&$RMSE~[10^{-3}]$ &							
			& 10	& 3	& 10	 & 4 
			\\\hline\hline
		\end{tabular}
	\end{table} 
	\par In Fig.~\ref{fig:distrib_12h}, the initial (at the start of collisions) and the final (after 11.5~h) longitudinal bunch profiles, as observed in the LHC (left) and as calculated by SIRE (right) for an initially $q$-Gaussian simulated profile, are denoted by blue and red stars, respectively. They are plotted in logarithmic scale  and they are fitted with the	Gaussian (dashed line) and the $q$-Gaussian (solid line) functions. The reduction of the bunch population with time due to burn-off and the extra (on top of IBS) transverse emittance blow-up observed in the machine, are taken into account for the simulation. The transverse distributions are assumed to be Gaussian, since at collisions the shape of their tails is not clear due to diffraction. The fitting results are presented in Table~\ref{tab:fitparams_12hFT}. Even if there seems to be no change at the tails of the simulated distribution, in reality the profiles become more Gaussian. Within 11.5~h at collisions, the rms beam size of the measured bunch profile and of the corresponding tracked distribution is reduced by  21\% and by 19\%, respectively. This shows a very good agreement between the experimental data and the simulations performed with SIRE. 
	\par
\begin{figure}[h]
		\centering
		\includegraphics[width=0.41\textwidth]{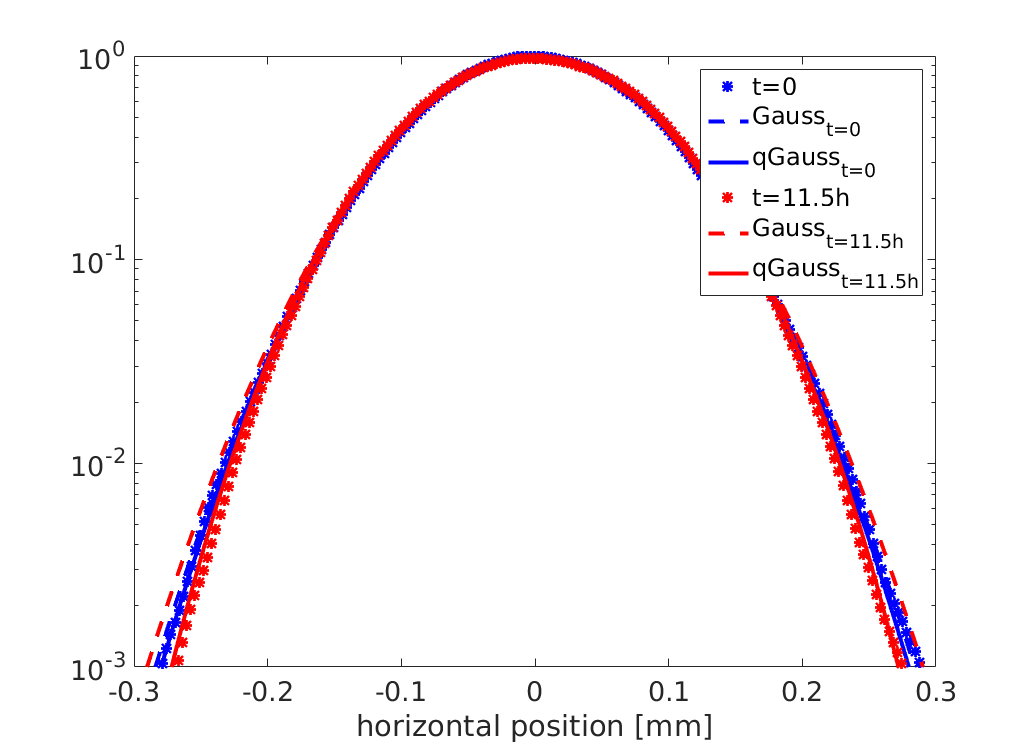}
		\includegraphics[width=0.41\textwidth]{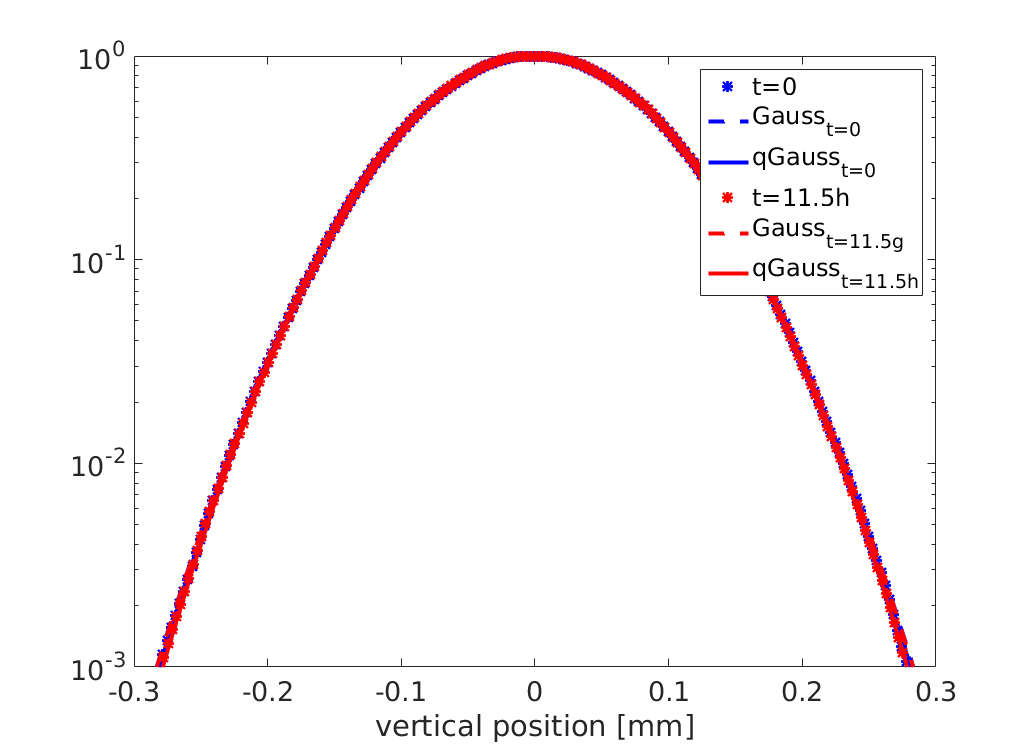}
		\captionsetup{justification=raggedright,singlelinecheck=false}
		\caption{The initial (at the start of collisions) and the final (after
			11.5 h) horizontal (left) and vertical (right) bunch profiles as calculated by the SIRE, in logarithmic scale, are denoted by blue and red stars, respectively. They are fitted with the Gaussian (dashed line) and the q-Gaussian (solid line) functions.}
		\label{fig:distrib_trans_12h}
	\end{figure}

 Figure~\ref{fig:distrib_trans_12h} shows in logarithmic scale the initial (blue stars) and the final (red stars) horizontal (left) and vertical (right) bunch profiles as calculated by SIRE, fitted with the Gaussian (dashed line) and the $q$-Gaussian (solid line) functions. As can be seen in Table~\ref{tab:fitparams_trans_12hFT}, the simulations showed no change in the transverse beam sizes and that is because the extra (on top of IBS) transverse emittance blow-up is included. The effect of IBS together with the extra blow-up assumed, widens the core of the horizontal bunch in such a way that the $q$ parameter is decreased by around 10\% within these 11.5~h.  Since IBS is negligible in the vertical plane, the fact that the vertical bunch profile remains Gaussian indicates that the interplay between the synchrotron radiation damping and the extra blow-up do not change the tails of the distribution. 
	\par
\begin{figure}[h!]
		\centering
		\includegraphics[width=0.35\textwidth]{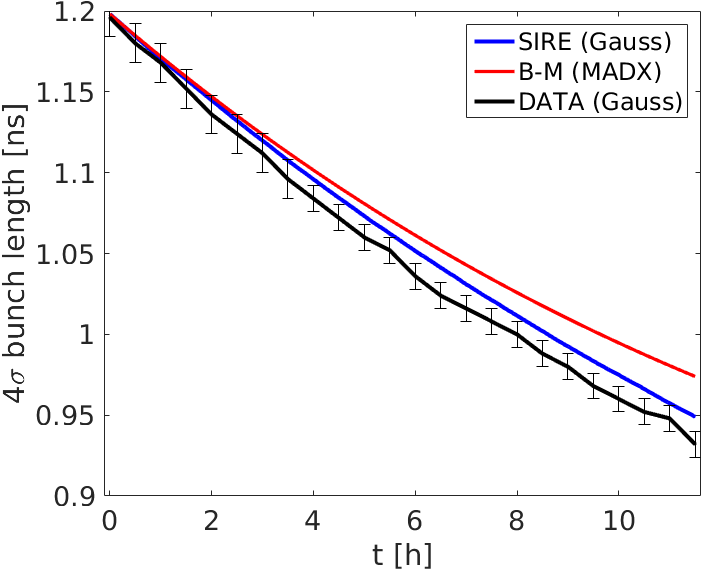}
		~~~~~	
		\includegraphics[width=0.35\textwidth]{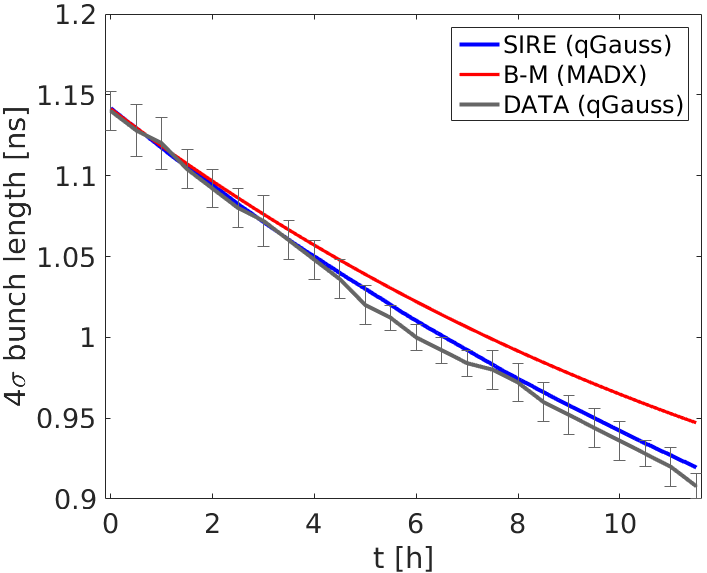}
		\captionsetup{justification=raggedright,singlelinecheck=false}
		\caption{The bunch length ($4\sigma$) evolution during several hours
			in collisions, as computed by the SIRE code (blue), the B-M
			analytical formalism (red) and as measured by the longitudinal
			profile monitors when assuming a Gaussian (left) and a q-Gaussian (right) distribution. }
		\label{fig:evol_12h}
	\end{figure}

  The $4\sigma$-bunch length evolution when assuming Gaussian (left) and $q$-Gaussian (right) initial distributions is shown in Figure~\ref{fig:evol_12h}. The blue line corresponds to the SIRE calculations and the red line to the results given by the IBS module of MAD-X~\cite{ref:ibsmadx} which is based on the analytical formulation of B-M and always assumes Gaussian distributions. The bunch length evolution, together with the two standard deviation error-bars, when fitting the data with the Gaussian and the q-Gaussian functions is represented by a black and a grey line, respectively.  The bunch length values differ for the two distribution functions used due to the fact that,  for a light tailed distribution the rms value is overestimated by fitting a Gaussian.  When assuming a Gaussian distribution, the bunch length evolution calculated by the code gets closer to the measured data. For the $q$-Gaussian case the agreement between data and simulations is excellent. This is a remarkable result, taking into account the fact that no assumptions are being made in the simulations apart from identical initial conditions with respect to the experimental ones.  In agreement with the results presented in the previous section, the divergence between the SIRE and the MAD-X for longer time-spans is something to be expected since the distribution shape in SIRE is updated, while in MAD-X it remains Gaussian.   
	
	\begin{table}[h!]
		\caption{Fitting results for the initial (at the start of collisions) and the final (after
			11.5 h) transverse bunch distributions shown in Fig.~\ref{fig:distrib_trans_12h}, as was calculated using the SIRE code.}
		\label{tab:fitparams_trans_12hFT}
		\centering
		\begin{tabular}{lcccccc}\hline\hline
			\multirow{2}{*}{\textbf{Fit Parameters}} & \multicolumn{2}{c}{ }		& \multicolumn{2}{c}{\textbf{Horizontal distribution}}	& \multicolumn{2}{c}{\textbf{Vertical distribution}}  		\\
			& &	&  $Initial$     	& $Final$ 	&  $Initial$     	& $Final$  \\\hline
			
			\multirow{3}{*}{q-Gaussian}	&$\sigma_{rms}\pm10^{-4}~[mm]$ &
			& $0.076$	& $0.076$	& $0.076$	& $0.076$ 	\\
			&$q\pm dq$ &	
			& $0.990\pm0.004$ & $0.893\pm0.005$ & $0.992\pm0.003$	& $0.983\pm0.003$ \\
			&$RMSE~[10^{-3}]$ &	
			& 3	& 3	& 3	 & 3 
			\\\hline\hline
		\end{tabular}
	\end{table}
	
\par  
	At collisions, the divergence between the luminosity model~\cite{ref:model} and the measured luminosity by the experiments becomes more pronounced as time evolves~\cite{ref:upmodel}. Actually, the predicted luminosity by the model is always larger compared to the measured one by the end of collisions. As calculated by SIRE, the weight of the horizontal bunch profile tails is decreased in time (see Table~\ref{tab:fitparams_trans_12hFT}) and as explained in Section~\ref{lumi}, for lighter tails and a constant beam size, the luminosity is expected to become lower. It is then clear that by taking into account the luminosity change due to the variation of the transverse distribution tails, the model predictions can be significantly improved.

\section{Summary and outlook}
\par In the LHC, the interplay between a series of effects  can lead to distributions with non-Gaussian tails. 
Since the rms value of a distribution can be underestimated or overestimated by using a simple Gaussian function, the use of appropriate fitting functions to accurately estimate the beam size and the behavior of the tails is necessary. The impact of non-Gaussian distribution shapes on the estimated luminosity is discussed. 
One of the next steps is to improve the luminosity model, that is currently based on Gaussian distributions, by taking into account the actual shape of the bunch profiles. In this way, it is possible to get more accurate luminosity predictions. Already, for the operational scenario of the High Luminosity LHC upgrade~\cite{ref:hllhc2}, a non-Gaussian bunch length estimation is being considered. 
\par  The way IBS and radiation effects act depends on the shape of the bunch profiles. Aiming to quantify the impact of the distribution's shape on the emittance evolution, a multi-particle tracking code called SIRE, is used. The benchmarking of the B-M analytical formalism with SIRE showed a very good agreement for the first couple of hours at the injection (450~GeV) and collision (6.5~TeV) energies of the LHC, even if they make use of different approaches to calculate the IBS effect. The differences observed for longer time-spans are expected, since in SIRE the particle distributions are updated, while MAD-X always assumes Gaussian distributions. The results obtained from the simulations encourage the idea of using the code for tracking distributions coming from experimental data, in order to study the impact of the distribution’s shape on the evolution of the bunch characteristics. After the comparison with experimental data, the fact that SIRE takes into account the change of the particle distribution showed that it is a very useful tool for estimating the actual bunch parameters evolution in the machine. The contribution of effects such as betatron coupling, noise and electron-cloud, to the emittance growth are planned to be included in the simulation code, in order to complement the existing semi-analytical emittance evolution model.

\appendix

\section{The q-Gaussian distribution function}
\label{qGaussian}

\begin{figure}[h]
	\centering
	\includegraphics[width=0.35\textwidth]{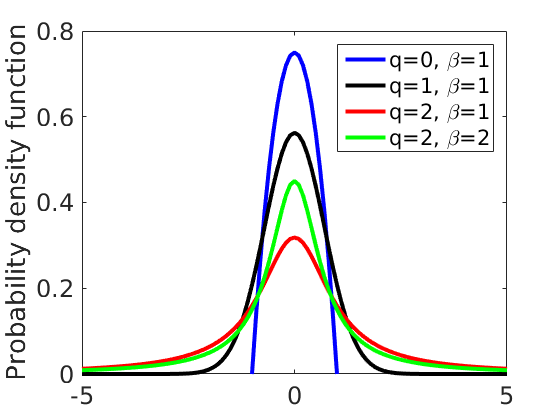}
	\caption{The q-Gaussian distribution function for different $q$ and $\beta^{qG}$ values.}
	\label{fig:qgaus}
\end{figure}

The q-Gaussian~\cite{ref:tsallis} which is used to describe more accurately bunch profiles with tails that differ from the ones of a normal distribution, has a probability density function given by: 
\begin{equation}
\label{eq:disfun_qGauss}
f(x) = {\sqrt{\beta^{qG}} \over C_q} e_q(-\beta^{qG} x^2)\;.
\end{equation}
The q-exponential function is given by:
\begin{equation}
\label{eq:fun_qGauss}
e_q(x)=
\begin{array}{cc}
\Bigg\{ & 
\begin{array}{l}
\exp(x)\:   ~~~~~~~~~~~~~~~,~\mathrm{if}~~~ q=1  \\
\left(1+ \left( 1-q\right) x\right)^{1 \over 1-q}\;,~\mathrm{if}~~~q\ne 1~\mathrm{and}~ \left(1+ \left( 1-q\right) x\right)>0 \\
0~~~~~~~~~~~~~~~~~~~~~~~,~\mathrm{if}~~~ q\ne 1~\mathrm{and}~ \left(1+ \left( 1-q\right) x\right)\le 0
\end{array}
\end{array}\:.
\end{equation}
The parameter $q$ describes the weight of the tails, in the sense that the larger its value, the heavier the tails become, as presented in Fig.~\ref{fig:qgaus}. In the limit of $q\rightarrow 1$, the normal distribution is obtained.
The distribution is characterized as ``light'' tailed when $q<1$ and as ``heavy'' tailed when $q>1$.
The normalization factor $C_q$ differs for specific ranges of the $q$ parameter and it is written as:
\begin{equation}
\label{eq:cq}
C_q=
\begin{array}{cc}
\Bigg\{ & 
\begin{array}{l}
\frac{2\sqrt{\pi}}{(3-q)\sqrt{1-q}} \frac{\Gamma\left({1 \over 1-q}\right) }{\Gamma\left({3-q \over 2(1-q)}\right)}\:,   ~~~\mathrm{for}~~~ -\infty<q<1  \\
\sqrt{\pi}  ~~~~~~~~~~~~~~~~~~~~~,~~~\mathrm{for}~~~ q=1 \\
\frac{\sqrt{\pi}}{ \sqrt{q-1}} \frac{\Gamma \left({3-q \over 2(q-1)}\right)}{\Gamma\left({1 \over q-1}\right)}  ~~~~~~\;,~~~\mathrm{for}~~~ 1<q<3
\end{array}
\end{array}\:.
\end{equation}
The parameter $\beta^{qG}$ is a 
real positive number. As the normal distribution, the $q$-Gaussian is an even function taking its maximum at $x=0$, where
\begin{equation}
f(0) = {\sqrt{\beta^{qG}} \over C_q}\;.
\end{equation}
For a certain $q$ value, the higher is the value of $\beta^{qG}$, the larger is the maximum of the probability density function, as can be observed in Fig.~\ref{fig:qgaus}. 
The standard deviation which also differs for specific ranges of the $q$ parameter, is given by:
\begin{equation}
\label{eq:sigma_qGauss}
\sigma^{qG}=
\begin{array}{cc}
\Bigg\{ & 
\begin{array}{l}
\sqrt{\frac{1} {\beta^{qG}(5-3q)}}\:,~~\mathrm{for}~~~ q<5/3  \\
\infty \;\;\;\;\;\;\;\;\;\;\;\,\,,~~\mathrm{for}~~~5/3\leq q<2 \\
\mathrm{undefined\,,~~for}~~~ 2\leq q<3
\end{array}
\end{array}\:.
\end{equation}
\par In the heavy tail regimes, the distribution is equivalent to the Student's t-distribution with a direct mapping between $q$ and the degrees of freedom $\nu$ (Eq.~\eqref{eq:student}). Statistically the q-Gaussian is a scaled re-parametrization of the Student's t-distribution~\cite{ref:student} for which the parameter $\nu$ is constrained to be a positive integer related to the sample size. The advantage of the q-Gaussian function is that, by introducing the parameters $q$ and $\beta^{qG}$, a generalization of the Student's t-distribution to negative and non-integer degrees of freedom is possible, where:
\begin{equation}
\label{eq:student}
q=\frac{\nu+3} {\nu+1} ~~~\mathrm{with}~~~ \beta^{qG}=\frac{1}{3-q}\:.
\end{equation}

\section{Luminosity calculation for q-Gaussian density distribution functions}
\label{lumi_qGaussian}
\par Using Eq.~\eqref{eq:disfun_qGauss} as the probability density functions, the general luminosity formula in Eq.~\eqref{eq:lumi_int} is solved for q-Gaussian distributions in all planes.
For the two beams being identical, integrating firstly over $s$ and $s_0$:
\begin{equation}
I^{qG}_{s}={\int\limits}{\int_{-\infty}}^{\infty}
\rho_{1s}(s-s_0) \rho_{2s}(s+s_0)~ds ds_0\:, 
\end{equation}
and then, integrating over x and y:
\begin{equation}
I^{qG}_{xy}={\int\limits}{\int_{-\infty}}^{\infty}
\rho_{1x}(x) \rho_{1y}(y) \rho_{2x}(x) \rho_{2y}(y) ~dx dy={\int\limits}{\int_{-\infty}}^{\infty}
\rho_{x}(x)^2 \rho_{y}(y)^2~dx dy\:,
\end{equation}
 keeping in mind that for $w=x, y, s$ it is:
\begin{equation}
\label{eq:xrange}
\begin{aligned}
w&\in \left[  \pm\dfrac{1}{\sqrt{\beta^{qG}(1-q)}}  \right]\:,~\mathrm{for}~ -\infty<q_w<1\\
w&\in (-\infty,\infty)\:,~~~~~~~~~~\mathrm{for}~~ 1\leq q_w<3\;, 
\end{aligned}
\end{equation}
the solutions of these integrals are found to be:
\begin{equation}
\label{eq:lumi_int_s}
I^{qG}_{s}=1\:, 
\end{equation}
and
\begin{equation}
\label{eq:lumi_int_xy}
I^{qG}_{x,y}=
\begin{array}{cc}
\Bigg\{ & 
\begin{array}{l}
\dfrac{\beta^{qG}_{x,y}}{C_{q_{x,y}}^2} \dfrac{\sqrt{\pi}\Gamma\left({-3+q_{x,y} \over -1+q_{x,y}}\right)}{\sqrt{\beta^{qG}_{x,y}(1-q_{x,y})}\Gamma\left({3q_{x,y}-7 \over 2(-1+q_{x,y})}\right)}\:,   ~~~\mathrm{for}~ -\infty<q_{x,y}<1  \\
\dfrac{\beta^{qG}_{x,y}}{C_{q_{x,y}}^2} \dfrac{\sqrt{\pi}\Gamma\left({-q_{x,y}+5 \over 2(-1+q_{x,y})}\right)}{\sqrt{\beta^{qG}_{x,y}(-1+q_{x,y})}\Gamma\left({2\over -1+q_{x,y}}\right)} ~\;,~~~\mathrm{for}~~~~~ 1\leq q_{x,y}<3
\end{array}
\end{array}\:,
\end{equation}
for $I^{qG}_{x}I^{qG}_{y}=I^{qG}_{xy}$ and, for $\beta^{qG}_{xy}$ and $C_{q_{xy}}$ being the beta parameters and the normalization factors in the transverse plane.
After some simplifications, using also Eq.~\eqref{eq:cq} and Eq.~\eqref{eq:sigma_qGauss}, it is found that the luminosity for q-Gaussian distribution functions depends on the ${\cal {I}}_{x,y}^{qG}$ (see~Eq.\eqref{eq:lumi_qG_trans}) which are defined as:
\begin{equation}
{\cal {I}}_{x,y}^{qG}=
\begin{array}{cc}
\Bigg\{ & 
\begin{array}{l}
\dfrac{(2+1/k)^2}{2\sqrt{3+2k}}\dfrac{\Gamma\left(1+2k\right)\Gamma\left(1/2+k\right)^2}{\Gamma\left(3/2+2k\right)\Gamma\left(k\right)^2}\:~~,   ~~~\mathrm{for}~ -\infty<q_{x,y}<1  
\\
\dfrac{2}{\sqrt{-(3+2k)}}\dfrac{\Gamma\left(-1/2-2k\right)\Gamma\left(-k\right)^2}{\Gamma\left(-2k\right)\Gamma\left(-1/2-k\right)^2}\:,~~~\mathrm{for}~~~~~~ 1\leq q_{x,y}<\dfrac{5}{3}
\end{array}
\end{array}\:,
\end{equation}
for $k=\dfrac{1}{1-q_{x,y}}$. As for the Gaussian case (Eq.~\eqref{eq:lumi_gauss2}), the luminosity for q-Gaussian beams colliding head-on does not depend on the longitudinal beam size.
\begin{figure}[h]
	\begin{center}
		\includegraphics[width=0.33\textwidth]{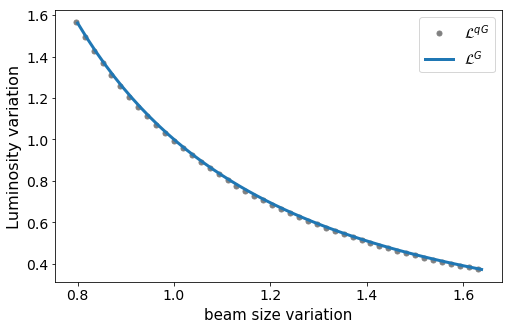}
		\captionsetup{justification=raggedright,singlelinecheck=false}
		\caption{Luminosity variation with respect to the transverse beam size variation, for the q-Gaussian (${\cal {L}}^{qG}$) with $q=1$ (i.e. normal distribution shape) and the Gaussian case (${\cal {L}}^G$), for the same beam parameters.}
		\label{fig:lumi_qconst}	
	\end{center}
\end{figure}
\par In Fig.~\ref{fig:lumi_qconst}, the variation of the luminosity is plotted with respect to the transverse beam size for the $q$-Gaussian case (${\cal {L}}^{qG}$) with $q=1$ (i.e. normal distribution shape) and the Gaussian case (${\cal {L}}^G$). Basically, the transverse beam sizes in Eq.~\eqref{eq:lumi_gauss2} and Eq.~\eqref{eq:lumi_qG_trans} are being varied equivalently and the resulted luminosity changes are found using these two equations.  The excellent agreement demonstrates that in the limit of $q\rightarrow1$, the luminosity estimation for $q$-Gaussian distributions (given in Eq.~\eqref{eq:lumi_qG_trans}) allows to obtain the exact same result as for Gaussian distributions. By keeping the $q$ parameter constant (here $q=1$) and varying the beam size, the $\beta^{qG}$ parameter also varies.

\bibliography{mybib}	
\end{document}